\documentclass[usenatbib]{mnras}

\usepackage{newtxtext,newtxmath}
\usepackage{textgreek}

\usepackage{ae,aecompl}

\usepackage{graphicx}	
\usepackage{amsmath}	
\usepackage{multicol}
\usepackage{hyperref}
\usepackage{ulem, soul}

\title[Clustering of Lockman Hole]{A study on the Clustering Properties of Radio-Selected sources in the Lockman Hole Region at 325 MHz}

\author[A Mazumder et al.]{
Aishrila Mazumder,$^{1}$\thanks{E-mail: aishri0208@gmail.com}
Arnab Chakraborty,$^{1,2,3}$
Abhirup Datta$^{1}$
\\
$^{1}$Department of Astronomy, Astrophysics and Space Engineering, Indian Institute of Technology Indore, Indore 453552, India\\
$^{2}$Department of Physics, McGill University, 3600 rue University, Montreal, QC H3A 2T8, Canada\\
$^{3}$McGill Space Institute, McGill University, 3550 rue University, Montreal, QC H3A 2A7, Canada
}
\date{Accepted XXX. Received YYY; in original form ZZZ}

\pubyear{2015}

\begin{document}
\label{firstpage}
\pagerange{\pageref{firstpage}--\pageref{lastpage}}
\maketitle

\begin{abstract}
 
Studying the spatial distribution of extragalactic source populations is vital in understanding the matter distribution in the Universe. It also enables understanding the cosmological evolution of dark matter density fields and the relationship between dark matter and luminous matter. Clustering studies are also required for EoR foreground studies since it affects the relevant angular scales. This paper investigates the angular and spatial clustering properties and the bias parameter of radio-selected sources in the Lockman Hole field at 325 MHz. The data probes sources with fluxes $\gtrsim$0.3 mJy within a radius of 1.8$^\circ$ around the phase center of a $6^\circ \times 6^\circ$ mosaic. Based on their radio luminosity, the sources are classified into Active Galactic Nuclei (AGNs) and Star-Forming Galaxies (SFGs). Clustering and bias parameters are determined for the combined populations and the classified sources. The spatial correlation length and the bias of AGNs are greater than SFGs- indicating that more massive haloes host the former. This study is the first reported estimate of the clustering property of sources at 325 MHz, intermediate between the preexisting studies at high and low-frequency bands. It also probes a well studied deep field at an unexplored frequency with moderate depth and area. Clustering studies require such observations along different lines of sight, with various fields and data sets across frequencies to avoid cosmic variance and systematics. Thus, an extragalactic deep field has been studied in this work to contribute to this knowledge. 



\end{abstract}




\section{Introduction}

Observations of the extragalactic sky at radio frequencies are essential for the study of both large-scale structures (LSS) and different populations of sources present in the Universe. The initial research on LSS using clustering was performed with the reporting of slight clustering signals from nearby sources \citep{seldner1981, Shaver1989}. With the advent of large-area surveys like FIRST (Faint Images of the Radio Sky at Twenty-Centimeters, \citealt{FIRST}) and NVSS (NRAO VLA Sky Survey, \citealt{Condon1998}), the studies became more precise due to the large number of sources detected in these surveys.

The extragalactic sky at radio frequencies is dominated by sources below mJy flux densities (at frequencies from MHz to a few GHz, see for example \citealt{simpson2006, mignano, seymour, Smolic2008, prandoni2018}). The source population can be divided into Active Galactic Nuclei (AGNs) and Star-Forming Galaxies (SFGs) \citep{Condon1989, Afonso_2005, simpson2006, bonzini, padovani2015, vernstrom2016}. The dominant sources at these fluxes are SFGs, AGNs of Fanaroff-Riley type I (FR I, \citealt{fanaroff_riley}), and radio-quiet quasars \citep{Padovani2016}. Emission mechanism dominating populations at low frequencies ($\lesssim$10 GHz) is synchrotron emission, modeled as a power law of the form $\textrm S_{\nu} \propto \nu^{-\alpha}$, where $\alpha$ is the spectral index. Study of the extragalactic population using synchrotron emission can help trace the evolution of the LSS in the Universe. It also helps to map their dependence on various astrophysical and cosmological parameters \citep{blake_wall, lindsay_first,hale_cosmos}. Radio continuum surveys, both wide and deep, help constrain the overall behavior of cosmological parameters and study their evolution and relation to the environment \citep{best, ineson, hardcastle2016, rivera, williams2018}. The clustering pattern of radio sources (AGNs and SFGs) can be studied to analyze the evolution of matter density distribution. Clustering measurements for these sources also provide a tool for tracing the underlying dark matter distribution \citep{PressSchechter1974, lacey_cole1993, lacey_cole_1994, sheth_tormen, Mo2010}. The distribution of radio sources derived from clustering is related to the matter power spectrum and thus provides insights for constraining cosmological parameters that define the Universe. 
The relationship of the various galaxy populations with the underlying dark matter distribution also helps assess the influence of the environment on their evolution. Clustering studies are also required for extragalactic foreground characterization for EoR and post-EoR science. Spatial clustering of extragalactic sources with flux density greater than the sub-mJy range (around$\sim$150 MHz) dominate fluctuations at angular scales of arcminute range. Thus, their modeling and removal allow one to detect fluctuation of the 21-cm signal on the relevant angular scales.

The definition of clustering is the probability excess above a certain random distribution (taken to be Poisson for astrophysical sources) of finding a galaxy within a certain scale of a randomly selected galaxy. This is known as the two-point correlation function \citep{Peebles1980}. The angular two-point correlation function has been studied in optical surveys like the 2dF Galaxy Redshift Survey \citep{Peacock2001, percival, norberg}, Sloan Digital Sky Survey \citep{einstein_bao,sdss_wang,sdss_simoni,Shi2016,sdss_10_bao} and the Dark Energy Survey \citep{des}. Optical surveys provide redshift information for sources either through photometry or spectroscopy. This information can be used to obtain the spatial correlation function and the bias parameter \citep{2df_spatial, Heinis2009, boss_tomography}. But for optical surveys, observations of a large fraction of the sky is expensive in terms of cost and time. Additionally, optical surveys suffer the limitation of being dust-obscured for high redshift sources. However, at radio wavelengths, the incoming radiation from these sources do not suffer dust attenuation and thus can be used as a mean to probe such high z sources \citep{Hao2011, cucciati, highz, Jarvis2016, saxena}. The highly sensitive radio telescopes like GMRT \citep{Swarup1991}, ASKAP \citep{askap}, LOFAR \citep{lofar} are also able to survey larger areas of the sky significantly faster. They are thus efficient for conducting large-area surveys in lesser time than the old systems while detecting lower flux densities. Therefore, radio surveys provide an efficient method for investigation of the clustering for the different AGN populations. Additionally, at low-frequencies ($\lesssim$ 1.4 GHz), synchrotron radiation from SFGs provide insight into their star-formation rates \citep{Bell2003, Jarvis2010, Davies2017, Delhaize2017, Gurkan2018}. These insights have lead to clustering studies of SFGs as well at low frequencies \citep{maglio2017, arnab2020}. Through clustering studies of radio sources, deep radio surveys help trace how the underlying dark matter distribution is traced by luminous matter distribution. In addition to this, the two-point correlation functions can also provide other information relevant for cosmology by fitting parameterized models to the data to obtain acceptable ranges of parameters. These include the bias parameter, dark energy equations of state, and $\Omega_{m}$ (total density of matter), to name a few \citep{Peebles1980, camera, raccanelli2012, planck2013, allison}.

Extensive observations at multiple frequencies can help understand the relationship of the various source populations with their host haloes and individual structures (stars) present. It has been inferred from clustering observations that AGNs are primarily hosted in more massive haloes than SFGs and are also more strongly clustered \citep{gilli, Donoso2014, maglio2017, hale_cosmos}. While AGNs are more clustered than SFGs, for the latter, the clustering appears to be dependent on the rate of star formation. SFGs with higher star formation rates are more clustered than the ones with a lower rate (since star formation rate is correlated to stellar mass, which in turn is strongly correlated to the mass of the host halo, see \citet{magnelli, Delvecchio2021, bonato2021lofar} and references therein). Studying the large-scale distribution of dark matter by studying the clustering pattern of luminous baryonic matter is vital for understanding structure formation.
From linear perturbation theory, galaxies are "biased" tracers of the underlying matter density field since they are mostly formed at the peak of the matter distribution \citep{Peebles1980}. Bias parameter (b) traces the relationship between overdensity of a tracer $\delta$ and the underlying dark matter overdensity ($\delta_{DM}$), given by $\delta = \textrm b\delta_{DM}$. The linear bias parameter is the ratio between the dark matter correlation function and the galaxy correlation function (\citet{Peebles1980, Kaiser1984, Bardeen1986}, also see \citet{DESJACQUES20181} for a recent review). Measurement of the bias parameter from radio surveys will allow measurements which probe the underlying cosmology governing the LSS, and probe dark energy, modified gravity, and non-Gaussianity of the primordial density fluctuations \citep{BLAKE20041063, CARILLI2004979, seljak, Raccanelli_2015, abdalla2015cosmology}.

Analysis of the clustering pattern for extragalactic sources is also important for observations targeting the 21-cm signal of neutral hydrogen (HI) from the early Universe. These weak signals from high redshifts have their observations hindered by many orders of magnitude brighter foregrounds -  namely diffuse galactic synchrotron emission \citep{Shaver1999}, free-free emission from both within the Galaxy as well as extragalactic sources \citep{Cooray2004}, faint radio-loud quasars \citep{DiMatteo2002} and extragalactic point sources \citep{dimatteo2004}. \citet{dimatteo2004} showed that spatial clustering of extragalactic sources with flux density $\gtrsim$0.1 mJy at 150 MHz (the equivalent flux density at 325MHz is $\sim$0.05 mJy) dominate fluctuations at angular scale $\theta \gtrsim$1$\arcmin$. Thus, their modeling and removal allow one to detect fluctuation of the 21-cm signal on relevant angular scales. So their statistical modeling is necessary to understand and quantify the effects of bright foregrounds. Many studies have modeled the extragalactic source counts as single power-law or smooth polynomial \citep{Intema16,franzen2019} and the spatial distribution of sources as Poissonian \citep{ali08} or having a simple power-law clustering. However, a Poisson distribution of foreground sources is very simplistic and may affect signal recovery for sensitive observations like those targeting the EoR signal \citep{ali08, Trott_2016}. Thus more observations are required for low-frequency estimates of the clustering pattern of compact sources.  

A number of studies have been done in recent years for observational determination of the clustering of radio selected sources (for instance \citet{Cress_1996, overzier2003, lindsay_first,maglio2017,hale_cosmos,Hale19,rana_tgss,arnab2020,lotss_clustering}). However, more such studies are required for modeling the influence of different processes on the formation and evolution of LSS in the Universe. The sample used for such analyses should not be limited to small deep fields, since the limited number of samples makes clustering studies of different populations (AGNs/ SFGs) sample variance limited. Studies on the statistics of the source distribution are also essential for understanding the matter distribution across space. Thus, observations using sensitive instruments are required to conduct more detailed studies. At 1.4 GHz and above, many clustering studies are present (for instance \citet{Cress_1996, overzier2003, lindsay_first, maglio2017, hale_cosmos, lh_clustering_1.4}); however there extensive studies at low frequencies (and wider areas) are still required. The TIFR GMRT Sky Survey (TGSS) \citep{Intema16} is a wide-area survey of the northern sky at 150 MHz. But the available catalog from the TGSS- Alternate Data Release (TGSS-ADR) suggests that the data is systematics limited. Thus it is unsuitable for large-scale clustering measurements \citep{Tiwari_2019}. The ongoing LOFAR Two-metre Sky Survey (LoTSS \citealt{lotss_dr1}) at a central frequency of 144 MHz is expected to have very high sensitivity and cover a very wide area and thus provide excellent data for studying source distribution statistics at low frequencies \citep{lotss_clustering}. However, to constrain cosmological parameters, consensus for the overall behavior of sources along different lines of sight and across frequencies is also required, and there data sets like the one analyzed here become important \citep{BLAKE20041063, CARILLI2004979, Norris2013}. Radio data has the advantage that even flux-limited samples contain high-z sources \citep{Dunlop1990}. Thus, using the entire radio band provides insights into physical processes driving the evolution of different galaxy populations and helps create a coherent picture of the matter distribution in the Universe. Therefore, studies at radio frequencies would help constrain the cosmology underlying structure formation and evolution. 

The recent study of the clustering of the ELAIS-N1 field centered at 400 MHz using uGMRT by \citet{arnab2020} was extremely sensitive, with an RMS ($\sigma_{400}$)\footnote{Unless otherwise stated, $\sigma_{frequency}$ is the RMS sensitivity at the quoted frequency throughout the text.} of 15 $\mu$Jy $\textrm beam^{-1}$. But the area covered was significantly smaller ($\sim$1.8 deg $^2$) than this work. This smaller field of view makes measurement of clustering properties on large angular scales impossible. Smaller areas also lead to smaller sample sizes for statistics, resulting in studies limited by cosmic variance. Another study of the HETDEX spring field at 144 MHz (using the data release 1  of LOFAR Two meter Sky Survey) by \citet{lotss_clustering} has a sky coverage of $\sim$350 square degree, but the mean $\sigma_{150}$ is $\sim$91 $\mu$Jy beam$^{-1}$. However, despite the sensitivity achieved in the survey, the analysis by \citet{lotss_clustering} is limited to flux densities above 2 mJy. Motivated by the requirement for a study in the intermediate range (in terms of flux density, area covered, and frequency), this work aims to quantify the clustering of the sources detected in the Lockman Hole field. The data analyzed here fall in the intermediate category, with a survey area $\sim$6 deg$^2$ with $\sigma_{325} \sim$ 50 $\mu$Jy beam $^{-1}$. It is thus ideal for clustering studies with a sizeable area of the sky covered (thus large angular scales can be probed) and moderately deep flux threshold (catalogue will have fluxes reliable to a lower value). Additionally, the Lockman Hole region has excellent optical coverage through surveys like SDSS and SWIRE; thus, associated redshift information is available to study spatial clustering and bias parameters. This frequency also has the additional advantage of having lesser systematics than the 150 MHz band while still being sensitive to the low-frequency characteristics of sources. New data releases for the LoTSS surveys promise greater sensitivity and source characterization over various deep fields targeted by these observations \citep{lotss2019, tasse2020}; all these observational data at multiple frequencies will put more precise constraints on the various parameters governing the structure formation and evolution. 

This work uses archival GMRT data at 325 MHz covering a field of view of $6^\circ \times 6^\circ$ through multiple pointings. In \citet{aishrila1}, data reduction procedure is described in detail. This work used the source catalogue obtained there for clustering analyses. However, the entire dataset could not be used due to limiting residual systematics at large angular scales. The clustering pattern and linear bias parameter are determined for the whole population and sub-populations, i.e., AGNs and SFGs, separately. The previous work by \citet{aishrila1} had determined the flux distribution of sources (i.e., differential source count) and characterized the spatial property and the angular power spectrum of the diffuse galactic synchrotron emission using the same data. 
 
This paper is arranged in the following manner: In section \ref{observation}, a brief outline of the radio data as well as various optical data used is discussed; the classification into source sub-populations is also using radio luminosity of sources is also shown. The following section, i.e., Section \ref{all_correlation} shows the clustering quantification - both in spatial and angular scales and calculation of linear bias for all the detected sources. Section \ref{sep_correlation} discusses the clustering property and bias for classified population, with a brief discussion on the choice of the field of view for this analysis discussed in Section \ref{discussion}. Finally, the paper is concluded in Section \ref{conclusion}.

For this work, the best fitting cosmological parameters obtained from the Planck 2018 data \citep{Planck2018I} has been used. The values are $\mathrm{\Omega_{M}}$ = 0.31, $\mathrm{\Omega_{\Lambda}}$ = 0.68, $\mathrm{\sigma_{8}}$ = 0.811, \& $H_{0}$ = 67.36 km s$^{-1}$ Mpc $^{-1}$. The spectral index used for scaling the flux densities between frequencies is taken as $\alpha$=0.8.

\section{Observations and Source Catalogues}
\label{observation}

\begin{figure*}

\includegraphics[width=5.0in,height=4.0in]{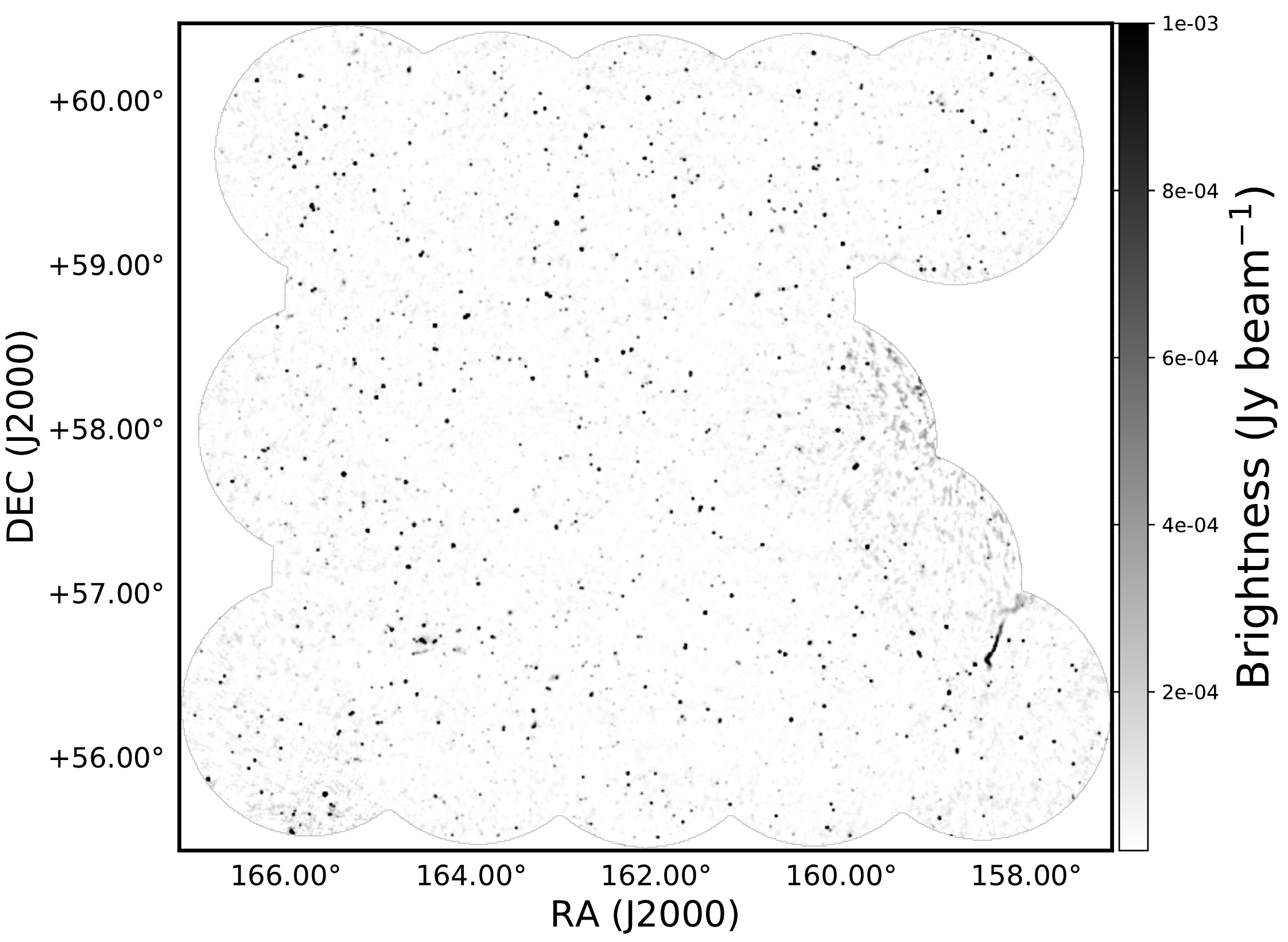}

 \caption{Primary beam corrected mosaic of the Lockman Hole region at 325MHz. The off source RMS at the center is $\sim$ 50 $\mu$Jy $\mathrm{beam}^{-1}$ and beam size is $ 9.0\arcsec \times 9.0\arcsec$. This image is a reproduction of Figure 1 of \citet{aishrila1}}.
\label{PB}
\end{figure*}

This work uses 325 MHz GMRT archival data of the Lockman Hole region. The details of the data reduction procedure have been described in \citet{aishrila1}, here it is discussed very briefly. The data were reduced using the SPAM pipeline \citep{Intema2009, Intema2014, Intema16}, which performs direction-independent as well as direction-dependent calibration techniques. The observation had 23 separate pointings , centered at ($\alpha_{2000}=10^{h}48^{m}00^{s},\delta_{2000}=58^{\circ}08'00\arcsec$), each of which was reduced separately. The final image is a $6^\circ \times 6^\circ$ mosaic having off-source RMS of 50$\mathrm{\mu Jy beam^{-1}}$ at the central frequency. Figure \ref{PB} shows the primary beam corrected final mosaic image of the observed region. This image was used to extract a  source catalogue using Python Blob Detection and Source Finder \footnote{\url{https://www.astron.nl/citt/pybdsf/}}(P{\tiny Y}BDSF, \citet{Mohan2015}) above a minimum flux density $\textrm S^{cut}_{325}$ 0.3mJy (i.e., above 6$\sigma_{325}$). A total of 6186 sources were detected and cataloged. The readers are referred to \citet{aishrila1} for details on catalogue creation and subsequent comparison with previous observations.

The redshift information for the sources are derived by matching with optical data from the Sloan Digital Sky Survey (SDSS)\footnote{\url{https://www.sdss.org/}}  and the Herschel Extragalactic Legacy Project (HELP) \footnote{\url{http://herschel.sussex.ac.uk/}}\textsuperscript{,}\footnote{\url{https://github.com/H-E-L-P}} \citep{help1}. The SDSS \citep{SDSSI, SDSSIII} has been mapping the northern sky in the form of optical images as well as optical and near-infrared spectroscopy since 1998. The latest data release (DR16) is from the fourth phase of the survey (SDSS-IV, \citet{Blanton2017}). It includes the results for various survey components like the extended Baryon Oscillation Spectroscopic Survey eBOSS, SPectroscopic identification of ERosita Sources SPIDERS, Apache Point Observatory Galaxy Evolution Experiment 2 APOGEE-2, etc. The surveys have measured redshifts of a few million galaxies and have also obtained the highest precision value of the Hubble parameter $H(z)$ to date \citep{sdss_hz}. An SQL query was run in the CasJobs \footnote{\url{https://skyserver.sdss.org/casjobs/}} server to obtain the optical data corresponding to the radio catalogue, and the catalogue thus obtained was used for further analysis. 

HELP has produced optical to near-infrared astronomical catalogs from 23 extragalactic fields, including the Lockman Hole field. The final catalogue consists of $\sim$170 million objects obtained from the positional cross-match with 51 surveys \citep{help1}. The performance of various templates and methods used for getting the photometric redshift is described in \citet{help2, duncan2018}. Each of the individual fields is provided separate database in the \textit{Herschel Database in Marseille} site \footnote{\url{https://hedam.lam.fr/HELP/}} where various products, field-wise and category wise are made available via "data management unit (DMU)". For the Lockman Hole field, the total area covered by various surveys is 22.41 square degrees with 1377139 photometric redshift objects. The Lockman Hole field is covered well in the  Spitzer Wide-area InfraRed Extragalactic Legacy Survey (SWIRE) with photometric redshifts obtained as discussed in \citet{robinson2008,robinson2012}. However, additional data from other survey catalogues like Isaac Newton Telescope - Wide Field Camera (INT-WFC, \citet{int}), Red Cluster Sequence Lensing Survey (RCSLenS, \citet{rcslens}) catalogues, Panoramic Survey Telescope and Rapid Response System 3pi Steradian Survey (PanSTARRS-3SS, \citet{panstarrs1}), Spitzer Adaptation of the Red-sequence Cluster Survey (SpARCS, \citet{sparcs}), UKIRT Infrared Deep Sky Survey - Deep Extragalactic Survey (UKIDSS-DXS, \citet{ukidss}), Spitzer Extragalactic Representative Volume Survey (SERVS, \citet{servs}) and UKIRT Hemisphere Survey (UHS, \citet{uhs}) resulted in more sources being detected and better photometric determination. The publicly available photometric catalogue for the Lockman Hole region was used to determine the redshift information for matched sources. The source catalogue derived from the 325 MHz observation is pre-processed, matched to add redshift information, and then further analysis is done. The following subsections describe these steps in detail.


\subsection{Merging multi-component sources}
The final map produced has a resolution of 9\arcsec. The source finder might resolve an extended source into multiple components for such high-resolution maps. Such sources are predominantly radio galaxies that have a core at the center and hotspots that extend along the direction of the jet(s) or at their ends; these structures may be classified as separate sources \citep{maglio98, prandoni2018, lofar_association, pink}. Using the NVSS catalogue, it has been shown in \citet{blake_wall_2002a} that large radio sources with unmarked components can significantly alter clustering measurements. Thus, for unbiased estimation of source clustering, such sources need to be identified and merged properly. A strong correlation between the angular extent of radio sources and their fluxes has been discovered by \citet{Oort1987}. The angular extent ($\theta$) of a source  is related to its flux density (S) by the $\theta$-S relation, $\theta \propto \sqrt{\textrm S}$. This relation was used to identify resolved components of multi-component sources in surveys like the FIRST survey \citep{maglio98}. 

Identification of multi-component sources in the Lockman Hole catalogue resolved as separate sources are made using two criteria. The maximum separation between pairs of sources (using the $\theta$-S relation) is given by $\mathrm{\theta_{max} = 20\sqrt{S_{total}}}$, where $\mathrm{S_{total}}$ is the summed flux of the source pairs \citep{Huynh_2005, prandoni2018,arnab2020}. Sources identified by the above criteria have been considered as the same source if their flux densities differ by less than a factor of 4 \citep{Huynh_2005}. 

\begin{figure}
\includegraphics[width=\columnwidth,height=3.0in]{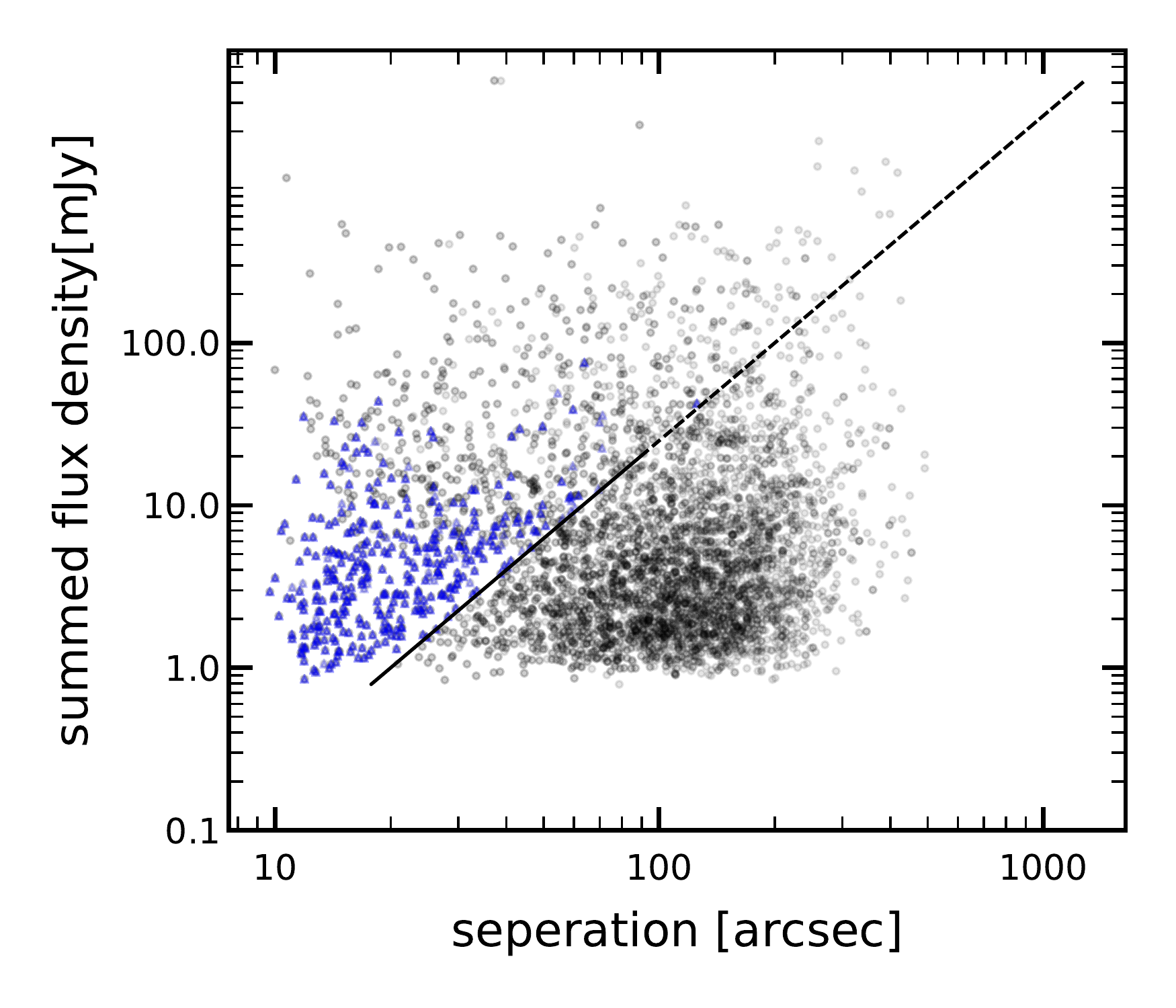}
 \caption{Sum of nearest neighbour flux densities from the 325 MHz catalogue as a function of separation between nearest neighbours in the catalogue are shown by light blue circles. Sources above the black doted line (having an angular separation less than $\mathrm{\theta_{max}}$) shown by blue triangles have their flux densities differing by a less than a factor of 4.}
\label{merge}
\end{figure}

Figure \ref{merge} shows the separation between the nearest neighbour pairs from the 325 MHz catalogue as a function of the separation between them. Above the black dotted line, the sources have separation less than $\mathrm{\theta_{max}}$ as mentioned above. Blue triangles are sources that have flux density differences less than a factor of 4. The two criteria mentioned gave a sample of 683 sources (out of 6186 total) to have two or more components. After merging multi-component sources and filtering out random associations, 5489 sources are obtained in the revised catalogue. The position of the merged sources are the flux weighted mean position for their components.

\subsection{Adding Redshift Information}
\label{redshift_dist}
As already mentioned, optical cross-identification have for the sources detected has been done using the HELP and SDSS catalogues.  A positional cross-match with 9\arcsec matching radius (which is the resolution for this observation) was used for optical cross-matching. Since the positional accuracy of the catalogue is better than 1\arcsec \citep{aishrila1}, a nearest neighbour search algorithm was used to cross-match sources with the optical catalogue with a search radius $r_s$. The rate of contamination expected due to proximity to optical sources is given by \citep{lindsay_first}:
\begin{equation*}
    P_{c} = \pi r_{s}^{2}\sigma_{opt}
\end{equation*}

where $\sigma_{opt}$ is the surface density of the optical catalogue. For surface density of 1.4$\times$10$^{4}$ deg$^{-2}$, a matching radius $r_s$ = 9\arcsec gives a contamination of \textless10\%. This radius was thus used to ensure valid optical identification of a large number of radio sources. 

\begin{figure}
\includegraphics[width=\columnwidth,height=3.0in]{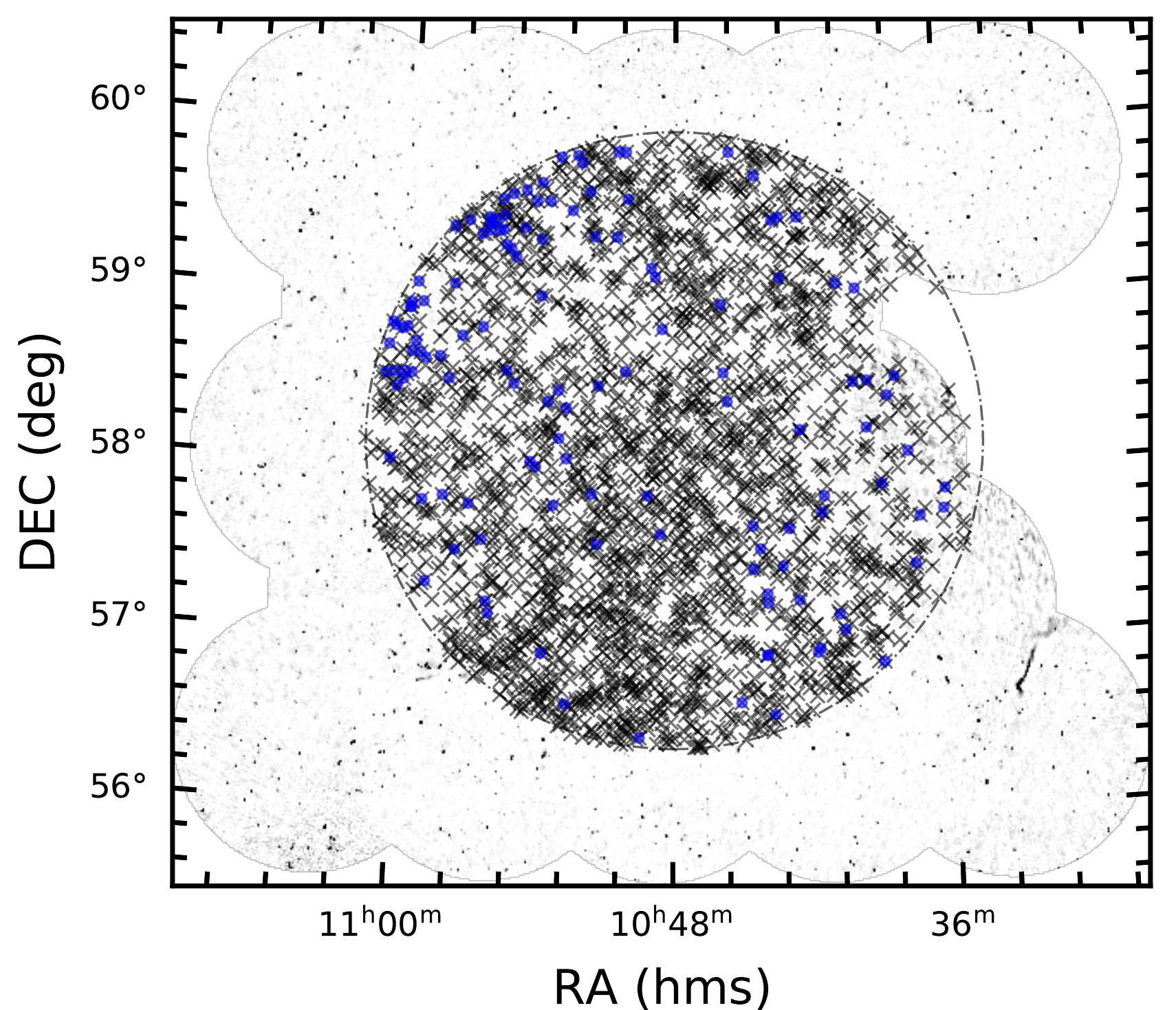}
 \caption{Sources with optical cross matches within 1.8$^\circ$ radius of phase center for the full mosaic. The area considered is represented by the black dot-dashed circle, and the sources in this region are represented by ``x" marks. The blue circles represent the sources without any redshift matches.}
\label{optical_ovreplot}
\end{figure}

Large FoVs are helpful for observational studies of like the one used here is useful for studying LSS, since the presence of a large number of sources provides statistically robust results and also reduces the effect of cosmic variance. Accordingly, the data used for this work had an Fov of $\rm 6^\circ \times 6^\circ$. However, cross-matching with optical catalogue produced matches with only 70\%  total sources over entire FoV, and 30\% sources remained unclassified. Further investigation also revealed the presence of unknown systematics, which resulted in excess correlation and deviation from power-law behavior at large angles. The most probable cause for such a deviation seems to be either the presence of many sources with no redshifts at the field edge or the presence of artifacts at large distances from the phase center. Analysis done by reducing the area of the field increased percentage of optical matches and reduced the observed deviation from the power-law nature. Thus, the cause of such a deviation has been attributed to the former one.

\begin{figure*}
\includegraphics[width=\columnwidth,height=3.0in]{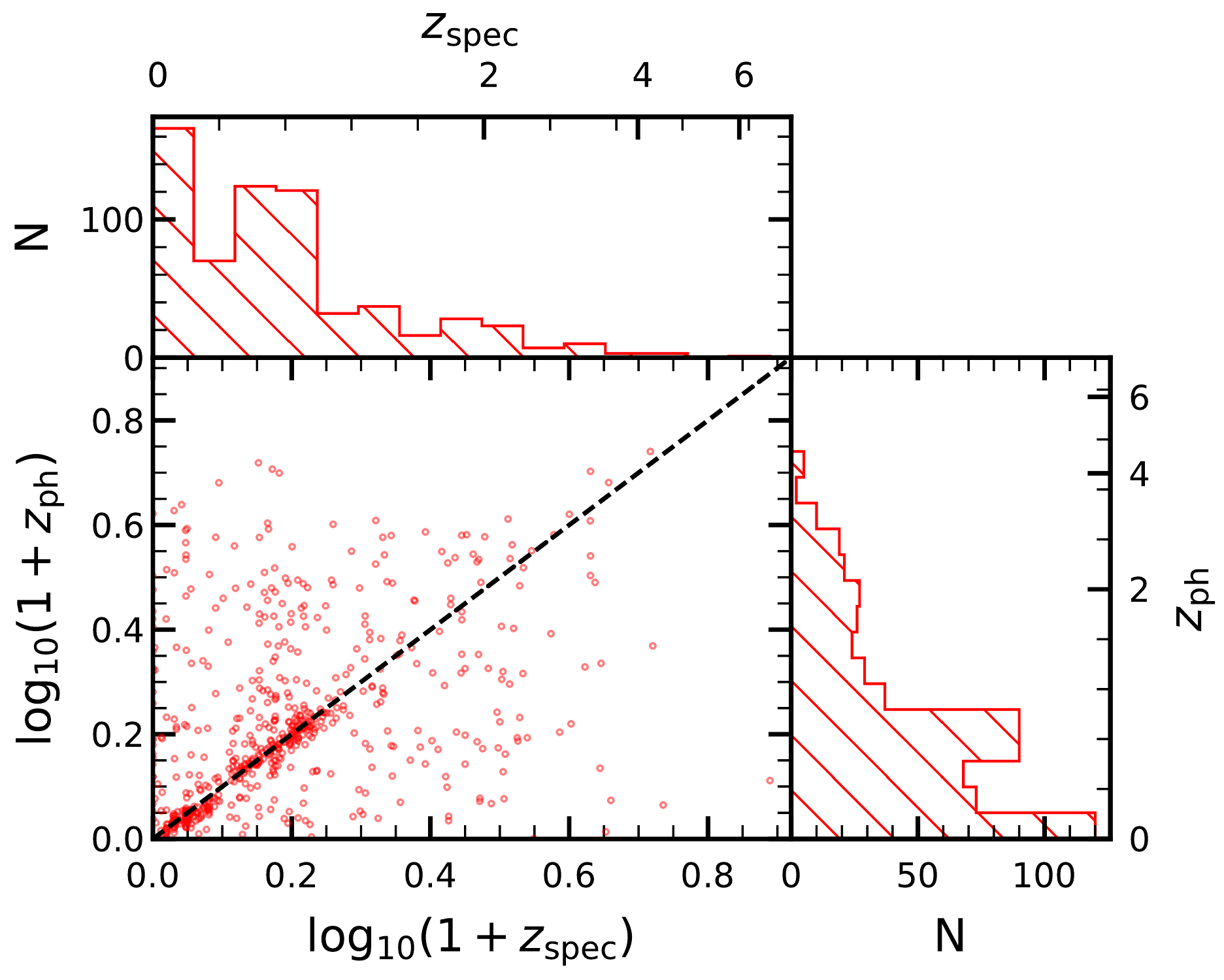}
\includegraphics[width=\columnwidth,height=2.8in]{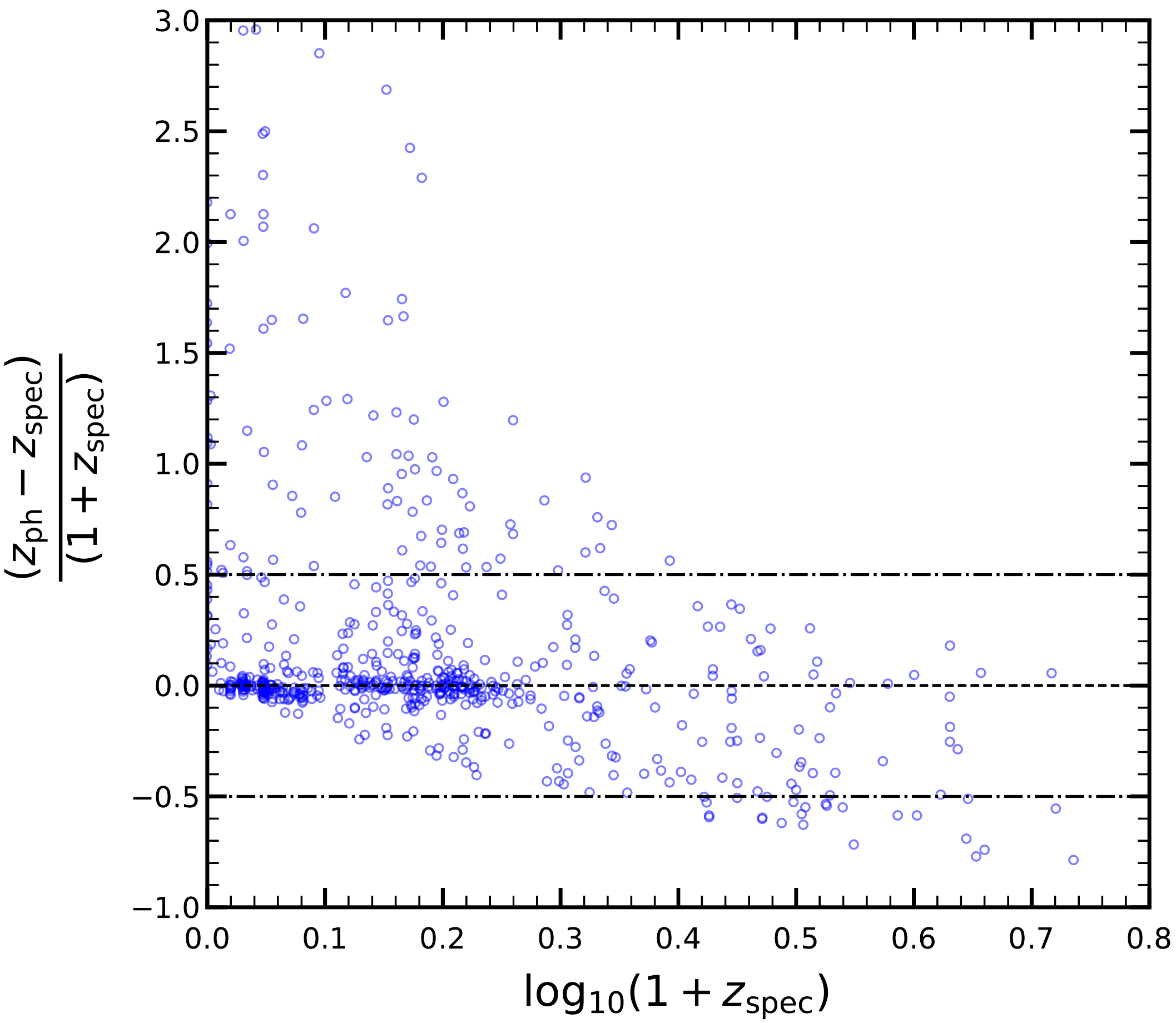}
 \caption{Comparison of photometric and spectroscopic redshifts for the sources with matches for both. \textbf{(left panel)} Photometric redshift plotted as a function of the spectroscopic redshift, with the black dashed line indicating equal values for both. The redshift distribution is shown via the histograms for photometric (right subplot) and spectroscopic (top subplot). \textbf{(right panel)} Distribution of $\mathrm{\frac{z_{phot}-z_{spec}}{1+z_{spec}}}$ plotted as a function of $\mathrm{z_{spec}}$. The black dashed line represents 0 deviation between the photometric and spectroscopic redshifts.}
\label{logz}
\end{figure*}

\begin{figure}
\includegraphics[width=\columnwidth,height=3.0in]{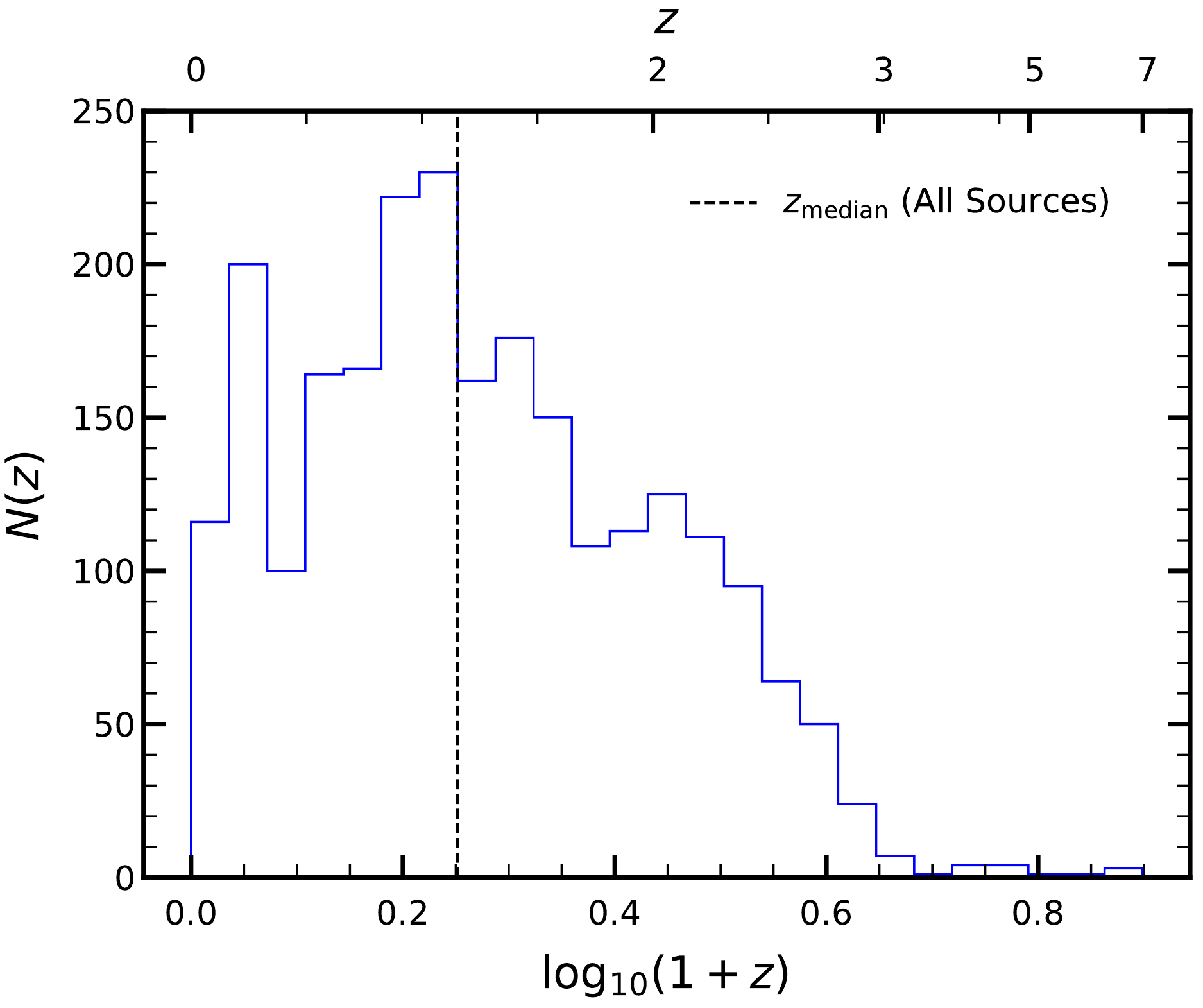}
\includegraphics[width=\columnwidth,height=3.0in]{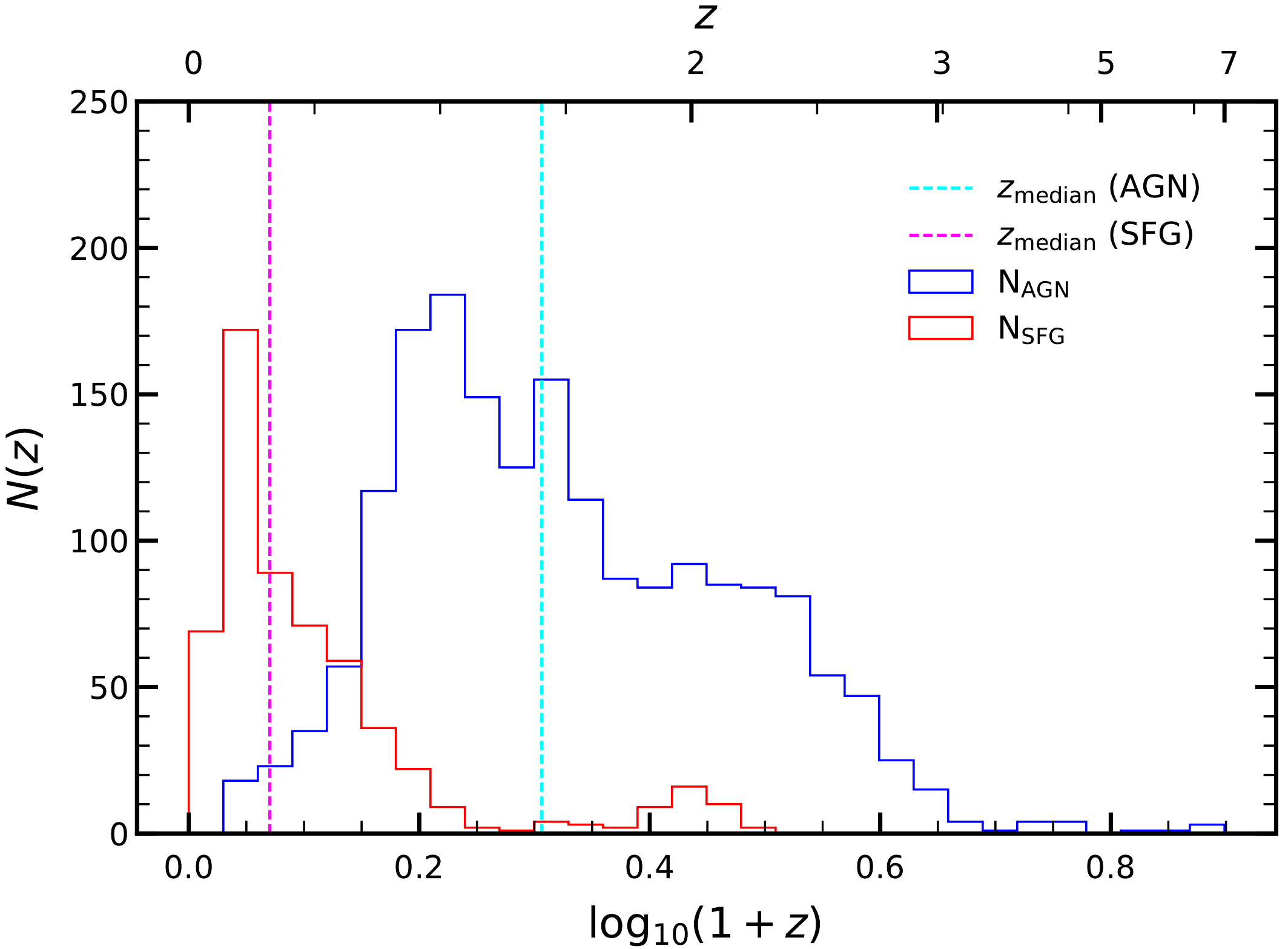}
 \caption{Redshift distribution (N($z$)) for sources with either photometric (HELP) or spectroscopic (SDSS) redshifts. The top panel shows N($z$) distribution for all sources with the black dashed line indicating the median redshift for all sources. The bottom shows the same for sources classified into AGNs (blue curve) and SFGs (red curve) using the radio luminosity criterion discussed in the text, with the cyan and magenta dashed lines indicating the respective median redshifts.}
\label{zhist}
\end{figure}

Hence, the clustering properties of sources at large angular scales are not reliable for this observation. Thus, the analysis was restricted to a smaller area of the Lockman Hole region around the phase center; large-scale clustering properties could not be estimated. Taking a cut-off with 1.8$^\circ$ radius around the phase center resulted in $\sim$95\% sources having an optical counter-part. Hence, it is expected that the unclassified sources present would not affect the signal significantly (a detailed discussion on the choice of the FoV cut-off is discussed in Section \ref{discussion}). This FoV cut-off yielded 2555 sources in the radio catalogue, out of which 2424 sources have optical matches within the aforementioned match radius. This is shown in Figure \ref{optical_ovreplot}, where the area considered is represented with black dot-dashed circle, and the "x" marks denote the sources in radio catalogue; the blue circles represent the sources without any optical cross-matches in either photometry or spectroscopy. A total of 2415 photometric and 664 spectroscopic matches were obtained after the cross-match with optical catalogues. Out of these, 650 sources had both photometric and spectroscopic detection. For such cases, the spectroscopic identifications were taken. Combined photometry and spectroscopic identifications were obtained for a total of 2424 sources, of which 27 sources were discarded from this analysis since they were nearby objects with 0 or negative redshifts \citep{lindsay_first}. The final sample thus had 2397 sources, which is $\sim$94\% of the total catalogued radio sources within 1.8$^\circ$ radius of the phase center. The redshift matching information for both the full and restricted catalogues have been summarised in Table \ref{z_summary}. The redshift information from the optical catalogues was incorporated for these sources and was used for further analysis. Figure \ref{logz} shows the distribution of redshifts for the sources detected in both HELP and SDSS. In the left panel, the photometric redshifts are plotted as a function of the spectroscopic redshifts. As can be seen, the two values are in reasonable agreement with each other for most cases. Additionally to check for the reliability of obtained photometric redshifts, following \citet{duncan2018}, the outlier fraction defined by $\mathrm{\frac{\mid z_{phot}-z_{spec}\mid }{1+z_{spec}}}$ \textgreater 0.2, is plotted as a function of the spectroscopic value (right panel of Figure \ref{logz}). For this work, the drastic outliers are the points with values \textgreater0.5. The fraction of outliers with drastically different values between photometric and spectroscopic redshifts is $\sim$10\%. While a detailed investigation is beyond the scope of this work, the outliers may be present due to the combination of uncertainties in the different surveys used in the HELP catalogue. As can be seen, the outlier fraction is not very drastic except for some cases; however, the reason for deviations in these sources is unknown. The median redshift for all the sources with redshift information comes out to be 0.78. The top panel of Figure \ref{zhist} shows the distribution N($z$) as a function of source redshift, with the black dashed line indicating median redshift. 

\begin{table*} 
\begin{center}
\caption{Summary of number of sources with redshift information}
\label{z_summary}
\begin{tabular}[\columnwidth]{lccccc}
\hline
\hline
Area & Number of sources & Redshift matches & Percentage of matches & AGNs & SFGs \\

\hline
$6^\circ \times 6^\circ$ & 5489 & 3628 & 66 & 2149 & 1479 \\
$3.6^\circ$ diameter around phase center & 2555 & 2397 & 95 & 1821 & 576\\
\hline
\end{tabular}
\end{center}
\end{table*}

\subsection{Classification using Radio Luminosity Function}
\label{classify}

The catalogued sources with optical counterparts were divided into AGNs and SFGs using their respective radio luminosities. Assuming pure luminosity evolution, the luminosity function evolves approximately as $\mathrm{(1+z)^{2.5}}$ and $\mathrm{(1+z)^{1.2}}$ for SFGs and AGNs respectively \citep{mcalpine2013}. The value for AGNs differ slightly from those of \citet{smolic_agn} and \citet{ocran_agn} for the COSMOS field at 3 GHz and ELAIS N1 field at 610 MHz respectively. However, they are consistent with those of \citet{prescott_gama} for the GAMA fields. The values for redshift evolution of SFGs also agree broadly for the GAMA fields \citep{prescott_gama} and the ELAIS N1 field \citep{ocran_sfg}.

It has been shown in \citet{maglio2014, maglio2016, maglio2017} that radio selected galaxies powered by AGNs dominate for radio powers beyond a radio power $\mathrm{P_{cross}}(z)$ which is related with the redshift $z$ as:
\begin{equation}
    \mathrm{log_{10}P_{cross}=log_{10}P_{0,cross}}+z
    \label{logp}
\end{equation}
upto $z \sim$1.8, with P (at 1.4 GHz) in W $\mathrm{Hz^{-1} sr^{-1}}$. In the local Universe, the value of $\mathrm{P_{cross}}$ is $\mathrm{10^{21.7}}$ (W $\mathrm{Hz^{-1} sr^{-1}}$), coinciding with the observed break in the radio luminosity functions of SFGs \citep{Maglio2002}, beyond which their luminosity functions decrease rapidly and the numbers are also reduced greatly. Thus contamination possibility between the two population of radio sources is very low using the radio luminosity based selection criterion \citep{maglio2014, maglio2017}.

The radio luminosity has been calculated for the sources from their flux as \citep{maglio2014}:
\begin{equation}
    \mathrm{P_{1.4GHZ} = 4 \pi S_{1.4GHz}D^{2}(1+z)^{3+\alpha}}
    \label{lum}
\end{equation}
where D is the angular diameter distance, and $\alpha$ is the spectral index of the sources in the catalogue. The individual spectral index for the sources was not used (since all sources do not have the measured values). The median value of 0.8 for $\alpha$ was derived by matching with high-frequency catalogues in \citet{aishrila1}. Since the probability of finding a large number of bright, flat-spectrum sources is very low \citep{maglio2017}, the median value of 0.8 was used to determine the luminosity functions of the sources in the Lockman Hole field detected here. 

Besides the radio luminosity criterion described above, there are several other methods to classify sources into AGNs and SFGs. X-ray luminosity can also be used to identify AGNs since it can directly probe their high energy emissions \citep{Szokoly_2004}. Color-color diagnostics from optical data (like IRAC) can also be used for identifying AGNs \citep{Donley_2012}. Classification can also be done using the q$_{24}$ parameter, which is the ratio of 24 $\mu$m flux density to the effective 1.4 GHz flux density \citep{bonzini}. Based on the resuts of \citet{mcalpine2013}, it was shown by \citet{maglio2014} that the radio luminosity function for SFGs fall of in a much steeper manner than AGNs for all redshifts, and this reduces the chances of contamination in the two samples. Additionally, these different multi-wavelength methods are not always consistent with each other, and a detailed investigation into any such discrepancy is beyond the scope of this work. Hence, only the radio luminosity criterion has been used for classification.

The sources with redshifts up to 1.8 were classified into AGNs and SFGs according to whether their luminosity is greater than or less than the threshold in Equation \ref{logp} (with $\mathrm{P_{cross}}$ determined using Equation \ref{lum}). At higher redshifts (i.e. \textgreater1.8), $\mathrm{P_{0,cross}}$ is fixed to 10$^{23.5}$ [$\mathrm{W Hz^{-1}sr^{-1}}$] \citep{mcalpine2013}. Of the 2397 sources, 1821 were classified as AGNs and 576 as SFGs using the radio luminosity criteria. The median redshifts for AGNs and SFGs are 1.02 and 0.2 respectively.

\section{Estimation of Correlation Function: Combined Sources}
\label{all_correlation}
\subsection{The Angular Correlation Function}

The angular two-point correlation function $w(\theta)$ is used to quantify clustering in the sky on angular scales. While several estimators have been proposed in literature (for a comparison of the different types of estimators see \citet{Kerscher_2000} and Appendix B. of \citet{lotss_clustering}), this work uses the LS estimator proposed by \citet{Landy1993}. It is defined as :
\begin{equation}
    w(\theta) = \mathrm{\frac{DD(\theta)-2DR(\theta)+RR(\theta)}{RR(\theta)}}
    \label{estimator_eq}
\end{equation}
Here DD($\theta$) and RR($\theta$) are the normalised average pair count for objects at separation $\theta$ in the original and random catalogues, respectively. Catalogue realizations generated by randomly distributing sources in the same field of view as the real observations have been used to calculate RR($\theta$). The LS estimator also includes the normalized cross-pair separation counts DR($\theta$) between original and random catalogue, which has the advantage of effectively reducing the large-scale uncertainty in the source density \citep{Landy1993, Hamilton1993, blake_wall_2002a, overzier2003}. The uncertainty in the determination of $w(\theta)$ is calculated using the bootstrap resampling method \citep{bootstrap}, where 100 bootstrap samples are generated to quote the 16th and 84th percentile errors in determination of $w(\theta)$. 

\subsubsection{Random Catalogue} 

The random catalogues generated should be such that any bias due to noise does not affect the obtained values of the correlation function. The noise across the entire $\mathrm{6^\circ \times 6^\circ}$ mosaic of the field is not uniform (see Figure 3 of \citet{aishrila1}). This can introduce a bias in estimating the angular two-point correlation function since the non-uniform noise leads to the non-detection of fainter sources in the regions with higher noise. 

P{\tiny Y}BDSF was used for obtaining the noise map of the image. Assuming the sources follow a flux distribution of the form dN/dS $\propto$ $S^{-1.6}$ \citep{intema2011, william2013}, random samples of 3000 sources were generated in the given flux range (with lower limit corresponding to 2 times the background RMS of the image) and assigned random positions to distribute them in the entire FoV. The sources constitute a mixture of 70\% unresolved sources and 30\% extended sources, which is roughly in the same ratio as the actual source catalogue \citep{aishrila1}. These were injected into the residual map, and using the same parameters in P{\tiny Y}BDSF as the ones used in the extraction of the original sources (see \citet{aishrila1}), the random catalogues were extracted. 100 such statistically independent realizations were used to reduce the associated statistical uncertainty. 

For clustering analysis of AGNs and SFGs, two sets of random catalogues were generated uing the publicly available catalogues for these source types from the T-RECS simulation \citep{trecs}. These catalogues have source flux densities provided at different frequencies between 150 MHz to 20 GHz. The flux densities at 300 MHz were considered for the randoms. They were scaled to 325 MHz using $\alpha=0.8$, and 2000 sources were randomly chosen within flux density limit for the radio catalogue of AGNs and SFGs. They were assigned random positions within the RA, Dec limits of the original catalogues and injected into the residual maps. Then using the same parameters for P{\tiny Y}BDSF as the original catalogue, the sources were recovered. 100 such realizations were done for AGNs and SFGs separately. The recovered random catalogues were used for further clustering analysis of the classified populations. It should also be mentioned here that the lower cut-off of flux density for the random catalogues was $\sim$0.1\,mJy, which is 2 times the background RMS. As already seen from \citet{aishrila1}, even a flux limit of 0.2\,mJy (4 times the background RMS) takes care of effects like the Eddington bias \citep{Eddington}. Thus 0.1\,mJy is taken as the limiting flux for both the combined and the the classified random catalogues. The final random samples for AGNs and SFGs consisted of a total of $\sim$120000 sources each, while for the combined sample, it was $\sim$200000. This is much higher than the number of sources in the radio catalogue. Thus, it does not dominate the errors. As has already been stated, the point and extended sources in the random catalogues (generated for the whole sample and the classified sources) are taken in the same ratio as that of the original radio catalogue. The drawback of this assumption is that there is a chance of underestimating extended sources in the random catalogue, which may lead to spurious clustering signals at smaller angular scales. However, since no evidence of any spurious signal is seen, taking point and extended sources in the same ratio as the original catalogue seem reasonable.

\subsection{Angular Clustering Pattern at 325 MHz}
\label{ang_325}

\begin{figure}
\includegraphics[width=\columnwidth,height=3.0in]{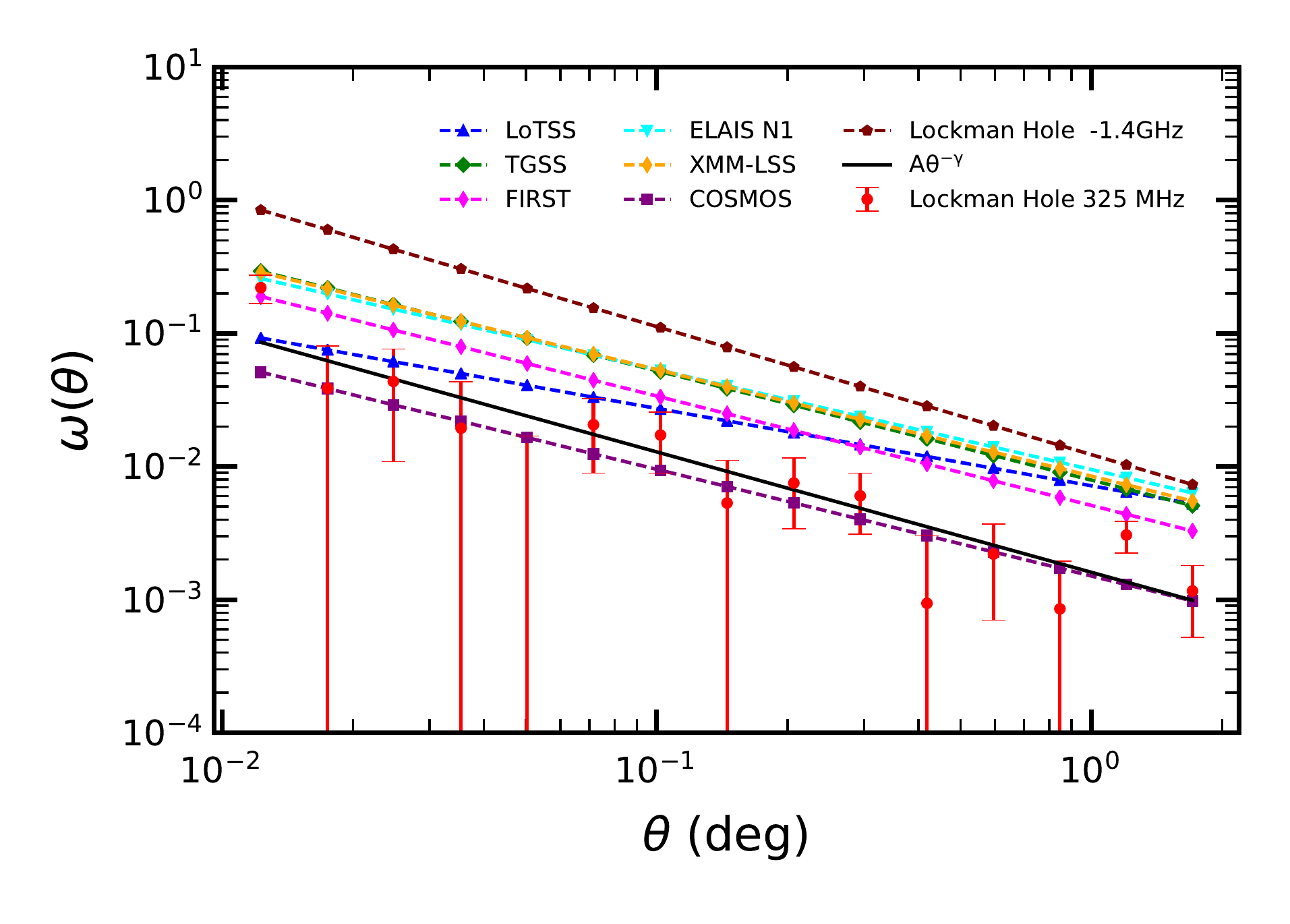}
 \caption{Angular correlation function for all sources in the Lockman Hole region at 325 MHz. The dashed black line is the best-fitting power law to $w(\theta)$. Comparison has been done with best fitted power laws obtained from previous studies of LoTSS (blue triangle, \citealt{lotss_clustering}), TGSS-ADR (green diamonds, \citealt{rana_tgss}), FIRST (magenta thin diamonds, \citealt{lindsay_first}), ELAIS-N1 (cyan inverted triangles, \citealt{arnab2020}), XMM-LSS (orange thin diamonds, \citealt{Hale19}), COSMOS (purple squares, \citealt{hale_cosmos}), \& Lockman Hole 1.4 GHz observation (maroon pentagons, \citealt{lh_clustering_1.4}). Different studies mentioned here have different flux density limits (see Table \ref{angular_table}).}
\label{angular_all}
\end{figure}

The angular correlation function of the sources detected in this observation is calculated using the publicly available code \texttt{TreeCorr} \footnote{\url{https://github.com/rmjarvis/TreeCorr}}\citep{treecorr}. The 325 MHz catalogue was divided into 15 equispaced logarithmic bins between $\theta \sim 36\arcsec$ to 2$^\circ$. The lower limit corresponds to the four times the PSF at 325 MHz, and the upper limit is the half-power beamwidth at this frequency. Figure \ref{angular_all} shows the angular correlation function of the 325 MHz in red circles; the error bars are estimated using the bootstrap method as discussed earlier. A power law of the form $w(\theta) = A\theta^{1-\gamma}$ is also fitted. The power law index, $\gamma$ is kept fixed at the theoretical value of 1.8. The parameter estimation for this fit is done using  Markov chain Monte Carlo
(MCMC) simulation by generating 10$^6$ data points by applying the Metropolis-Hastings algorithm in the $A$ parameter space. The first 10$^2$ samples have been removed from the generated chains to avoid the burn-in phase. From the sampled parameter space, $\chi^2$ is used to estimate the most likely values of the parameters. The best fit parameters are $\mathrm{log(A) = -2.73^{+0.11}_{-0.15}}$, with the error bars being the 1-$\sigma$ error bars from the 16th and 84th percentiles of the chain points.

\subsubsection{Comparison with previous Observations}

The best fit values obtained for parameters A and $\gamma$ of the 325 MHz catalogue have been compared with those for other observations at radio frequencies. The parameters obtained for different radio surveys, namely from \citet{lindsay_first, hale_cosmos, Hale19, rana_tgss, arnab2020, lh_clustering_1.4, lotss_clustering} have been summarised in Table \ref{angular_table}. The best-fit estimate of the slope $\gamma$ for the correlation function is found to be in reasonable agreement with the theoretical prediction of \citet{Peebles1980} and previous observations (for example see \citet{lh_clustering_1.4,lotss_clustering}).
The scaled flux limit at 325 MHz for the \citet{lh_clustering_1.4} catalogue at (originally 1.4 GHz, scaled using a spectral index of 0.8) is $\sim$0.4 mJy, very close to the flux limit for this work. However, their estimates are higher than all previous estimates (they particularly compare with \citep{maglio2017}), which they assign partly to the presence of sample variance. While the area probed by \citep{lh_clustering_1.4} is also included within the region this work probes, the area covered are different, the one covered here being larger.
This might be the reason for differences between the estimates in this work and \citep{lh_clustering_1.4}, despite both having similar flux density cut-offs. The clustering amplitude for this work is similar to \citet{hale_cosmos} at almost all the angular scales. One possible reason is that the flux limit for the study at 3 GHz was 5.5 times the 2.3 $\mu$Jy $\mathrm{beam}^{-1}$ limit corresponding to a flux of $\sim$0.1 mJy at 325 MHz, which is near the flux cut-off for this work (0.3 mJy), and thus can trace similar halo masses and hence clustering amplitudes.

\begin{table*} 
\begin{center}
\caption{Clustering Parameters for Observed Data. The columns indicate the name of the survey (Observation), observing frequency in MHz (Frequency), the flux density cut-off at the observing frequency (S$_\mathrm{cut,\nu}$), the equivalent 325 MHz flux-density (S$_\mathrm{cut,325}$), best fit clustering amplitude ($\mathrm{log_{10}(A)}$) and best fit power-law index ($\gamma$) respectively.}
\label{angular_table}
\begin{tabular}[\columnwidth]{lccccll}
\hline
\hline
Observation & Frequency  & S$_\mathrm{cut,\nu}^\dagger$ & S$_\mathrm{cut,325}^*$ & $\mathrm{log_{10}(A)}$ &  $\gamma$ & Reference \\
       &  (MHz)   & (mJy) & & &\\
 \hline
FIRST & 1400 &1.00 & 3.21 & -2.30$^\mathrm{^{+0.70}_{-0.90}}$  & 1.82$\pm$.02 & \citet{lindsay_first}\\
COSMOS & 3000 & 0.013 & 0.08& -2.83$^\mathrm{^{+0.10}_{-0.10}}$ & 1.80 & \citet{hale_cosmos} \\
XMM-LSS & 144 & 1.40 & 0.73 & -2.08$^\mathrm{^{+0.05}_{-0.04}}$ & 1.80 & \citet{Hale19} \\
TGSS-ADR & 150 & 50 & 26.9 &-2.11$^\mathrm{^{+0.30}_{-0.30}}$ & 1.82$\pm$.07 & \citet{rana_tgss}\\
ELAIS-N1 & 400 & 0.10 & 0.12 &-2.03$^\mathrm{^{+0.10}_{-0.08}}$ & 1.75$\pm$0.06 & \citet{arnab2020}\\
Lockman Hole & 1400 & 0.12 & 0.39 &-1.95$^\mathrm{^{+0.005}_{-0.005}}$ & 1.96$\pm$.15 & \citet{lh_clustering_1.4}\\
LoTSS & 144 & 2.00 & 1.04 & -2.29$^\mathrm{^{+0.6}_{-0.6}}$ & 1.74$\pm$.16 & \citet{lotss_clustering} \\
Lockman Hole & 325 & 0.30 & 0.30 & -2.73$\mathrm{^{+0.11}_{-0.15}}$ & 1.80 & This work\\
\hline
\hline
\end{tabular}
\end{center}
 \flushleft{$^\dagger$ S$_\mathrm{cut,\nu}$ is the flux density limit at the respective observing frequencies; $^*$ S$_\mathrm{cut,325}$ is the scaled flux density ($\alpha$=0.8) limit at 325 MHz\\
 }
\end{table*}

The clustering properties of the radio sources in the VLA-FIRST survey \citep{FIRST} has been reported in \citet{lindsay_first}, where  $\mathrm{log(A)}$ is -2.30$^\mathrm{^{+0.70}_{-0.90}}$. \citet{hale_cosmos} and \citet{Hale19} have reported $\mathrm{log(A)}$ value of -2.83 and -2.08 for the COSMOS and XMM-LSS fields, respectively, by fixing $\gamma$ at the theoretical value of 1.80. The clustering amplitude of the 150 MHz TGSS-ADR \citep{Intema16} has been shown by \citet{rana_tgss} for a large fraction of the sky and at different flux density cut-offs. In the recent deep surveys of the ELAIS-N1 field at 400 MHz \citep{arnab2020}, $\mathrm{log(A)}$ and the best fit power law index have the values -2.03$^\mathrm{^{+0.10}_{-0.08}}$ and 1.75$\pm$0.06 respectively. 

Comparison has also been made with the wide-area survey of LoTSS data release 1 \citep{lotss_clustering}. This study (with data obtained at a central frequency of 144 MHz) employed various masks on the data to obtain the angular clustering values. The survey covers a wider area, but the flux cut-off threshold is above 1 mJy for all of the masks due to systematic uncertainties. A wide range of angles, 0.1$^\circ \leq \theta \leq$ 32$^\circ$ was fixed to determine the angular clustering. Taking three different flux density limits- at 1, 2 and 4 mJy and different masks, the values of $\mathrm{log(A)}$ and power-law index were obtained(the fitting for the power-law form was done for 0.2$^\circ \leq \theta \leq$ 2$^\circ$). \citet{lotss_clustering} have applied various flux density cuts and masks to their sample for obtaining the angular clustering parameters. They have concluded that the flux density cut-off of 2 mJy provides the best estimate for the angular clustering parameters, and the same has been used here for comparison. Comparison of the present work with LoTSS 2 mJy flux cut shows that the values of $\mathrm{log(A)}$ agree well. The best fit power-law index is also consistent within 1$\sigma$ error bars. Hence, it is seen that the angular correlation function obtained in the present work gives values for the parameters $\mathrm{log(A)}$ and $\gamma$ consistent with those reported in previous surveys. Additionally, since this survey has both wider coverage than the recent EN1 data and a lower flux density threshold than the LoTSS data used by \citet{lotss_clustering}, it provides an intermediate data set along a different line of sight to probe cosmology.

\subsection{The Spatial Correlation Function at 325 MHz}

For known angular clustering $w(\theta)$, the spatial clustering of sources is quantified by the two-point correlation function $\xi(r)$. Using the Limber inversion \citep{Limber1953}, $\xi(r)$ can be estimated for known redshift distribution. Gravitational clustering causes the spatial clustering to vary with redshift, and thus a redshift dependent power-law spatial correlation function can be defined as \citep{Limber1953, overzier2003}:
 \begin{equation}
     \mathrm{\xi(r,z)} = \mathrm{(r_{0}/r)^{\gamma} (1+z)^{\gamma-(3+\epsilon)}}
     \label{spatial_limber}
 \end{equation}

where the clustering length r is in comoving units, $\epsilon$ specifies clustering models \citep{overzier2003} and $\mathrm{r_{0}}$ is the clustering length at z=0. For this work, comoving clustering model, in which the correlation function is unchanged in the comoving coordinate system and with $\epsilon$ = $\gamma$-3, is used. The comoving cluster size is constant. The correlation length is calculated using \citep{Peebles1980}:

\begin{equation}
    \mathrm{A} = \mathrm{r_{0}^{\gamma}H_{\gamma} (H_{0}/c)\frac{\int_{0}^{\infty}N^{2}(z)(1+z)^{\gamma-(3+\epsilon)}\chi^{1-\gamma}(z)E(z)dz}{[\int_{0}^{\infty}N(z)dz]^{2}}}
    \label{smooth_function}
\end{equation}

where $\mathrm{H_{\gamma} = \frac{\Gamma(\frac{1}{2})\Gamma(\frac{\gamma-1}{2})}{\Gamma(\frac{\gamma}{2})}}$, \\$\mathrm{E(z)=\sqrt{\Omega_{m,0}(1+z)^{3}+\Omega_{k,0}(1+z)^{2}+\Omega_{\Lambda,0}}}$ is the cosmological factor, N(z) is the redshift distribution of the sources and $\mathrm{\chi(z)}$ is the line of sight comoving distance. Equation \ref{smooth_function} can be used to estimate $\mathrm{r_{0}}$ using the angular clustering amplitude A and the redshift distribution shown in Figure \ref{zhist}. 

\begin{figure}
\includegraphics[width=\columnwidth,height=3.0in]{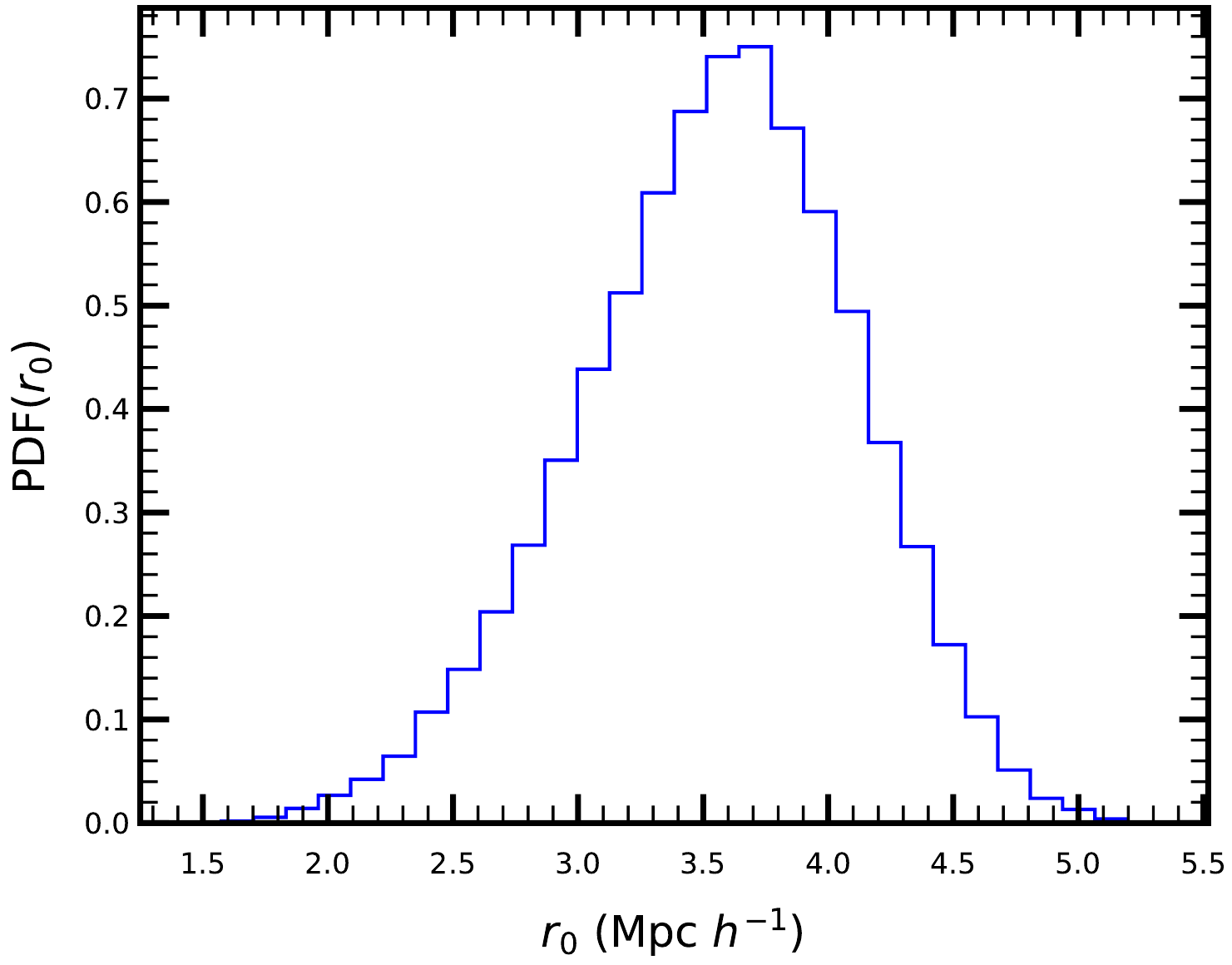}
 \caption{Probability distribution function of spatial clustering length (\rm $r_{0}$) for the entire sample at 325 MHz.}
\label{space_pdf}
\end{figure}

The theoretical value of 1.8 for $\gamma$, as predicted by \citet{Peebles1980} is consistent with the values across various surveys, as well as within 2 $\sigma$ of the current analysis (tabulated in Table \ref{angular_table}). Thus the theoretical value of $\gamma$, the distribution of A obtained from the MCMC distribution discussed in \ref{ang_325} and the combined redshift distribution distribution discussed in Section \ref{redshift_dist} are used to estimate the value of $\mathrm{r_{0}}$. Figure \ref{space_pdf} shows the probability distribution function (PDF) of the spatial clustering length. As already mentioned, the median redshift of the samples is $\sim$0.78, and at this redshift, the median value of $\mathrm{r_{0}}$ is 3.50$\mathrm{^{+0.50}_{-0.50}}$ Mpc h$^{-1}$, where the errors are the 16th and 84th percentile errors. 

\subsection{The Bias Parameter}

The bias parameter is used to quantify the relation between the clustering property of luminous sources and the underlying dark matter distribution. The ratio of the galaxy to the dark matter spatial correlation function is known as the scale-independent linear bias parameter $b(z)$ \citep{Kaiser1984, Bardeen1986, peacock}. For cosmological model with dark matter governed only by gravity,  following \citet{lindsay_first,hale_cosmos,arnab2020}, $b(z)$ is calculated as : 
\begin{equation}
    b(z) = \Bigg(\frac{r_{0}(z)}{8}\Bigg)^{\gamma/2}\frac{J_{2}^{1/2}}{\sigma_{8}D(z)/D(0)}
\end{equation}

where $J_{2}$ = 72/[(3-$\gamma$)(4-$\gamma$)(6-$\gamma$)2$^{\gamma}$], $D(z)$ is the linear growth factor, calculated from CMB and galaxy redshift information \citep{Eisenstein_1999}, and $\sigma_{8}^{2}$ is the amplitude of the linear power spectrum  on a comoving scale of 8 Mpc $\rm h^{-1}$. 

For this work, the bias parameter has been calculated using the median redshift value of the r$_\mathrm{0}$ distribution with the 16th and 84th percentile errors. The value of the bias parameter $b(z)$ at $z$=0.78 is 2.22$\mathrm{^{+0.33}_{-0.36}}$.

\section{Estimation of Correlation Function: AGNs and SFGs}
\label{sep_correlation}

\begin{figure*}
\includegraphics[width=\columnwidth,height=3.0in]{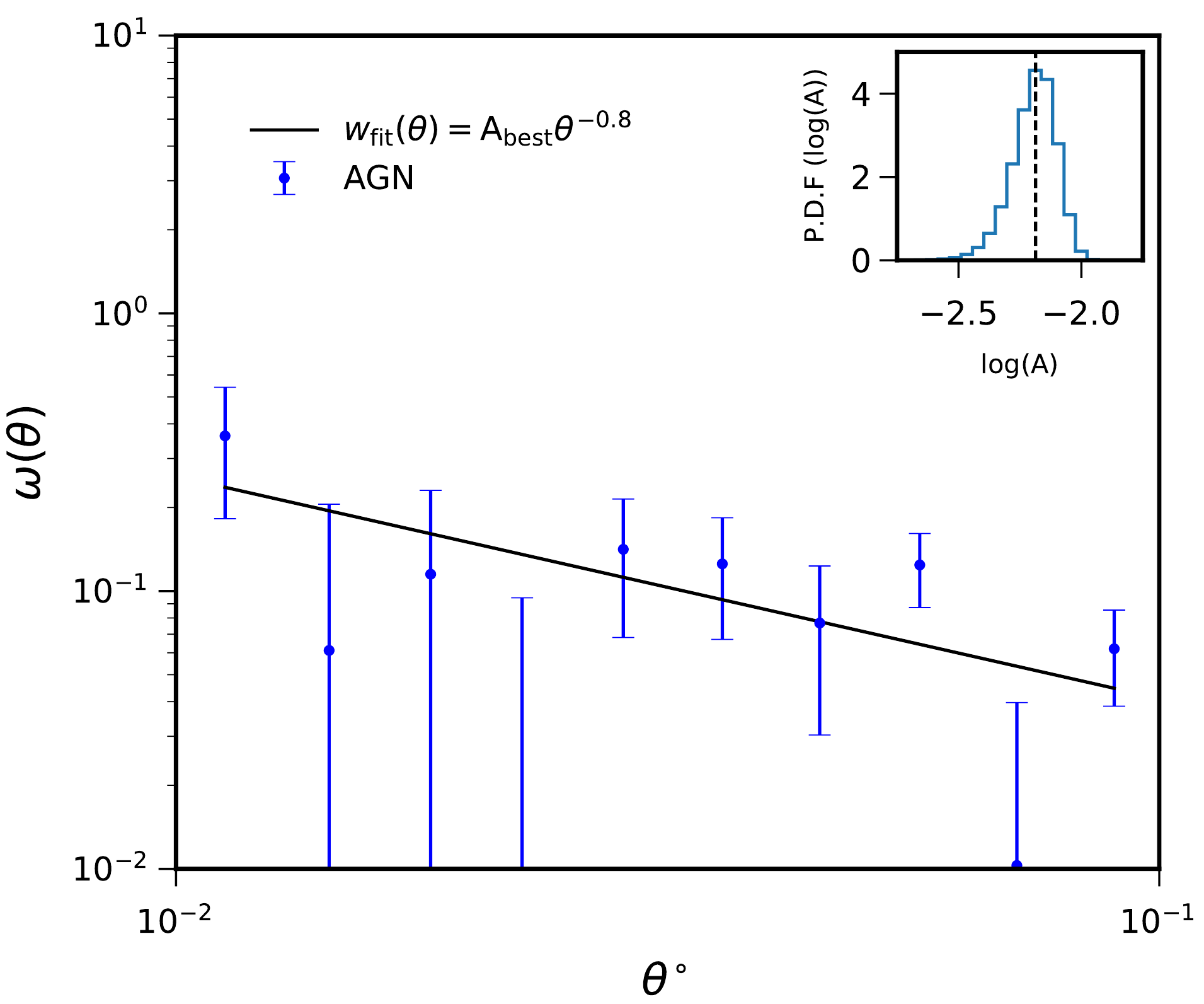}
\includegraphics[width=\columnwidth,height=3.0in]{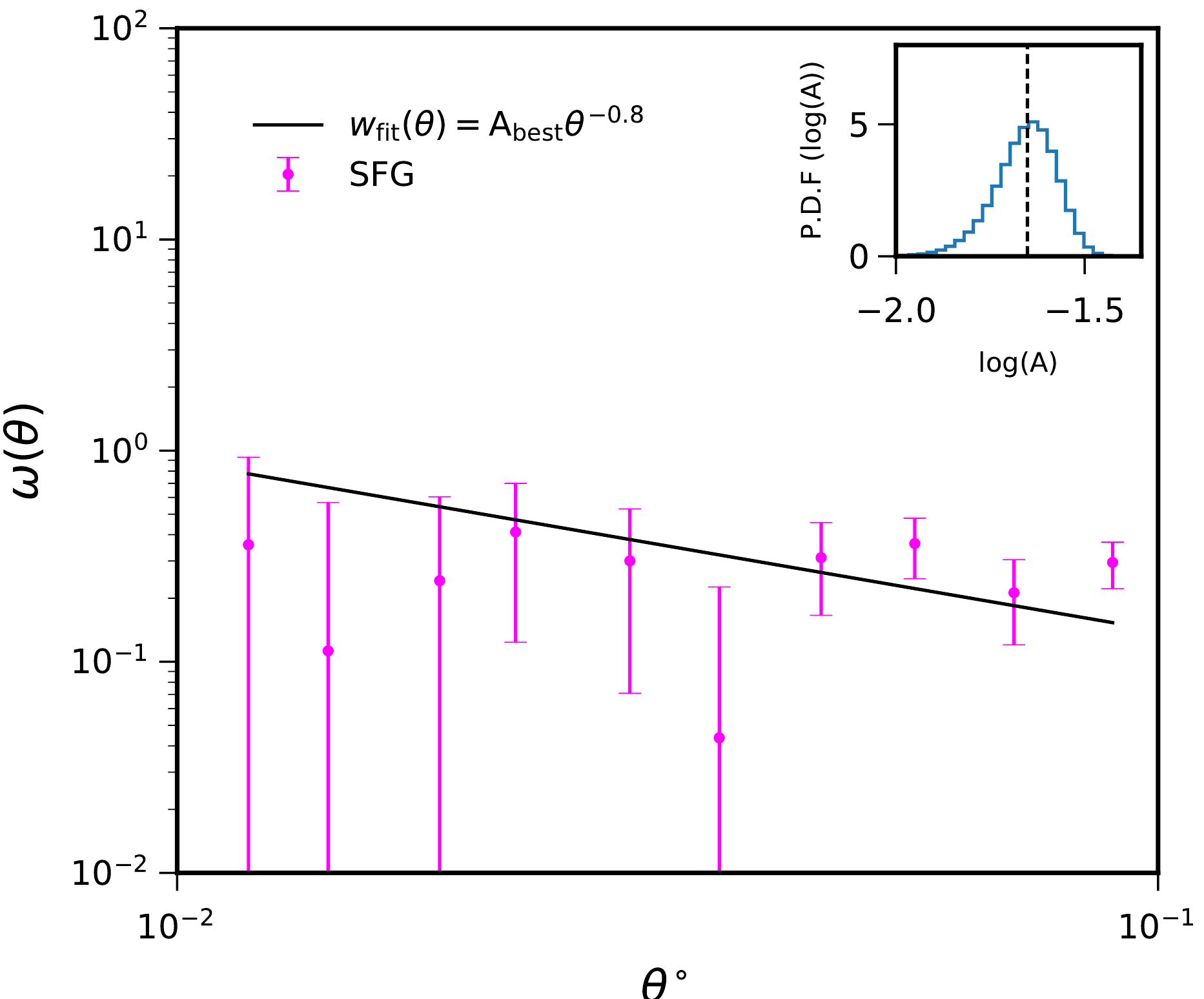}
 \caption{Angular correlation function for sources classified as AGNs (left panel) and SFGs (right panel). The slope of the best fit function $\gamma$ is fixed at 1.8 and the distribution of correlation amplitude ($A$) is shown in the inset of each panel.}
\label{angular_sep}
\end{figure*}

This section discusses the angular and spatial correlation scales and the bias parameter obtained for the two separate populations of sources (i.e., AGNs and SFGs). The obtained values are also compared with previously reported values using radio and other bands data. Following a similar procedure as that done for the entire population, initially, the angular clustering was calculated, and a power law of the form $A\theta^{1-\gamma}$ was fitted. The best value of clustering amplitude $A$ is determined, once again keeping $\gamma$ fixed at the theoretical value of 1.8 for both AGNs and SFGs populations. Figure \ref{angular_sep} shows the angular correlation function of AGNs (left panel) and SFGs(right panel). Using the MCMC simulations as discussed previously, the clustering amplitudes, $\mathrm{log(A)}$ have values $\textrm -2.18^{+0.20}_{-0.20}$ and $ \textrm -1.69^{+0.10}_{-0.10}$  respectively for AGNs and SFGs. The results of the fit and the subsequent values of clustering length and bias parameter obtained here and results from previous surveys in radio wavelengths are also tabulated in Table \ref{spatial_table}.  

The spatial clustering length and bias parameter b$_{z}$ for the AGNs with  $z_{median}$=1.02 are $8.30^{+0.96}_{-0.91}$ Mpc $\rm h^{-1}$ and $3.74^{+0.39}_{-0.36}$.
For SFGs with $z_{median}\approx$0.20, the values are $\mathrm{r_{0}}$= $3.22^{+0.34}_{-0.32}$ Mpc $\rm h^{-1}$ and b$_{z}$=$1.06^{+0.1}_{-0.1}$. 
It is seen that the spatial clustering length and consequently the bias factor for AGNs is more than SFGs, which implies that the latter are hosted by less massive haloes, in agreement with previous observations \citep{gilli, gilli07, Starikova_2012, Dolley_2014, maglio2017, hale_cosmos, arnab2020}. 

\subsection{Comparison with previous Observations}

\begin{table*} 
\begin{center}
\caption{Spatial Clustering Length and Bias Parameter from Different Observations. The columns are respectively name of the survey field, observing frequency in MHz, type of radio source (AGNs/SFGs), median redshift, angular clustering amplitude, spatial clustering length in Mpc h$\rm ^{-1}$ \& bias parameter value.}
\label{spatial_table}
\begin{tabular}[\columnwidth]{lccccccc}
\hline
\hline
Observation & Frequency & Source type & $\textrm z_{median}$ & $\mathrm{log_{10}(A)}$ & $\mathrm{r_{0}}$ & b$_{z_{median}}$ & Reference \\
&(MHz)&& &&(Mpc h$\rm ^{-1}$)&&\\
\hline
COSMOS & 3000 & AGNs & 0.70 & $-2.30^{+0.1}_{-0.1}$ & $6.9^{+0.60}_{-0.70}$ & $2.1^{+0.2}_{-0.2}$ & \citet{hale_cosmos}\\
&& AGNs & 1.24 & $-2.60^{+0.1}_{-0.1}$ & $9.6^{+0.70}_{-0.70}$ & $3.6^{+0.2}_{-0.2}$ &\\
&& AGNs & 1.77 & $-2.60^{+0.1}_{-0.1}$ & $7.3^{+0.90}_{-0.90}$ & $3.5^{+0.4}_{-0.4}$ &\\
&& SFG & 0.62 & $-2.60^{+0.1}_{-0.1}$ & $5.0^{+0.50}_{-0.60}$ & $1.5^{+0.1}_{-0.2}$ &\\
&& SFG & 1.07 & $-2.90^{+0.1}_{-0.1}$ & $6.1^{+0.60}_{-0.70}$ & $2.3^{+0.2}_{-0.2}$ &\\
&&&&&&&\\
VLA-COSMOS &1400 & AGNs & 1.25 & $-2.79^{+0.1}_{-0.1}$ & $7.84^{+1.75}_{-2.31}$ & - & \citet{maglio2017}\\
&& SFG & 0.50 & $-2.36^{+0.3}_{-0.3}$ & $5.46^{+1.12}_{-2.10}$ & - &\\
&&&&&&&\\
ELAIS N1 & 400 & AGNs & 0.91 & $-2.22^{+0.16}_{-0.16}$ & $7.30^{+1.4}_{-1.2}$ & $3.17^{+0.5}_{-0.5}$ & \citet{arnab2020}\\
&& SFG & 0.64 & $-2.16^{+0.05}_{-0.06}$ & $4.62^{+0.39}_{-0.40}$ & $1.65^{+0.14}_{-0.14}$ &\\
&&&&&&&\\
ELAIS N1 & 612 & AGNs & 0.85 & $-2.30^{+0.02}_{-0.03}$ & $6.0^{+1.5}_{-1.3}$ & $2.6^{+0.6}_{-0.5}$ & \citet{arnab2020}\\
&& SFG & 0.87 & $-2.19^{+0.01}_{-0.02}$ & $4.16^{+0.7}_{-0.8}$ & $1.59^{+0.2}_{-0.2}$ &\\
&&&&&&&\\
Lockman Hole & 325 & AGNs & 1.02 & $-2.18^{+0.20}_{-0.20}$ & $8.30^{+0.96}_{-0.91}$ & $3.74^{+0.39}_{-0.36}$ & This work\\
&& SFG & 0.20 & $-1.65^{+0.1}_{-0.1}$ & $3.22^{+0.34}_{-0.32}$ & $1.06^{+0.10}_{-0.10}$ &\\
\hline
\hline
\end{tabular}
\end{center}

\end{table*}

\begin{figure*}
\includegraphics[width=\columnwidth,height=3.0in]{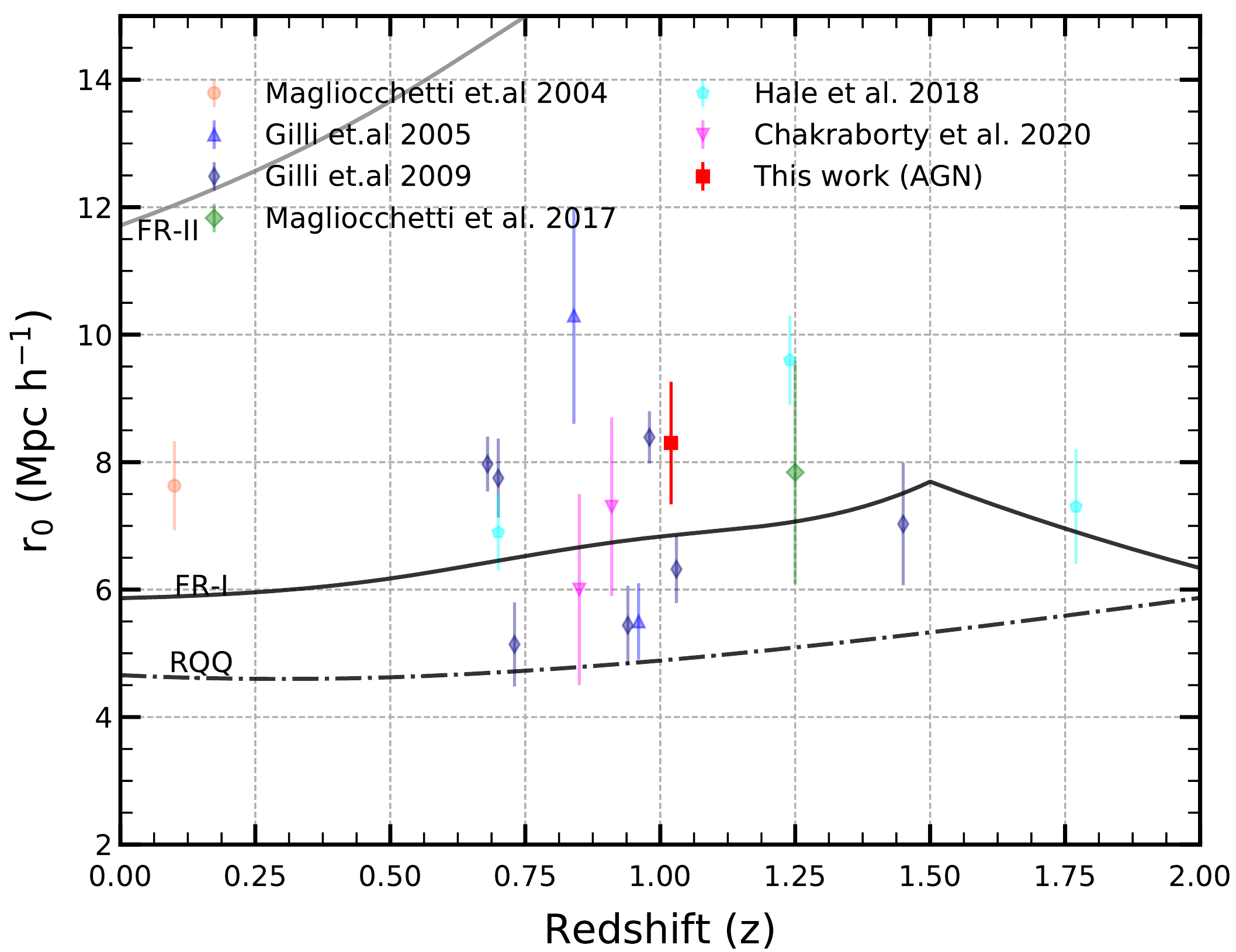}
\includegraphics[width=\columnwidth,height=3.0in]{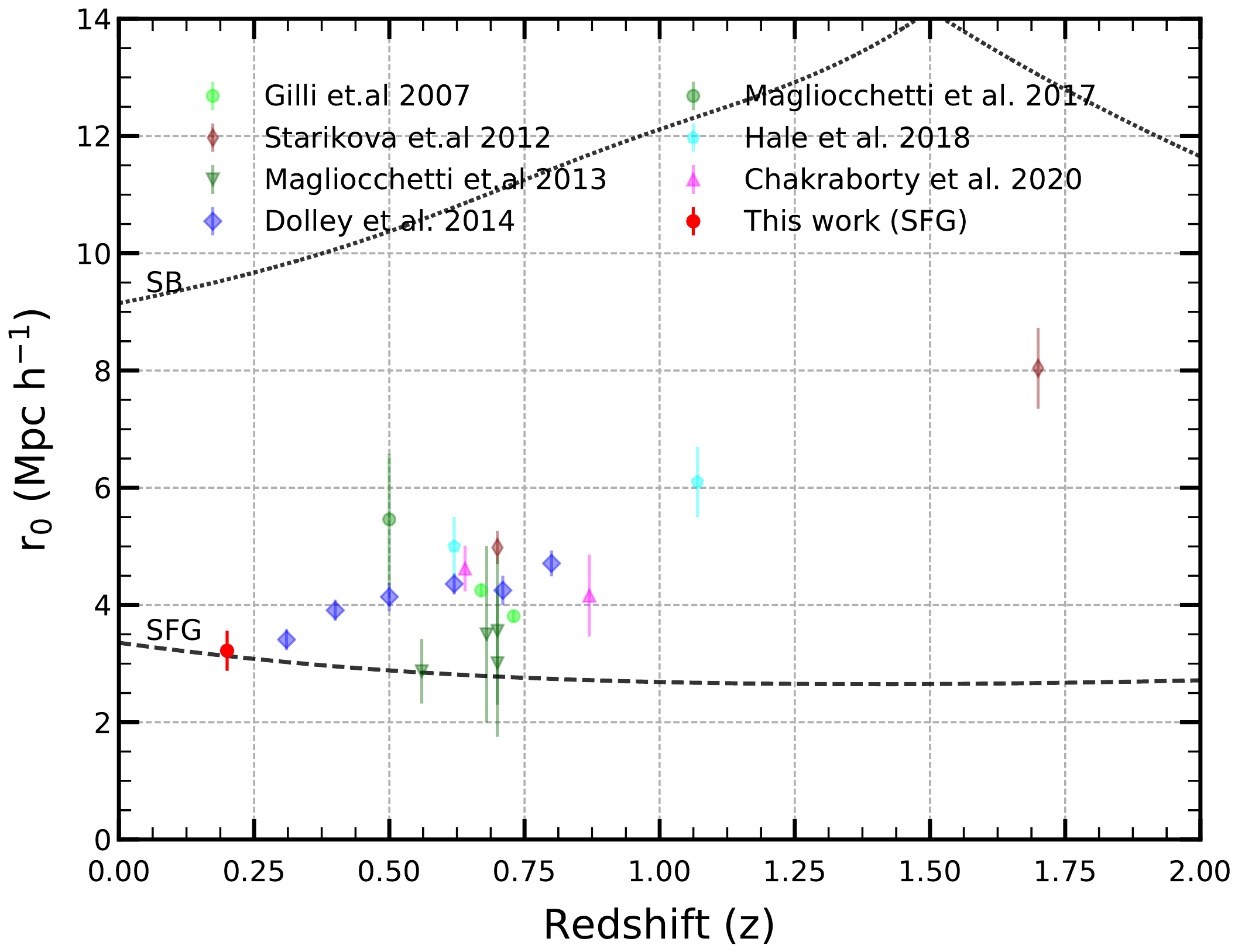}
 \caption{\textbf{(left panel)} Spatial correlation length ($\mathrm{r_{0}}$) for AGNs as a function of redshift (red square). For comparison, the values for the same obtained by \citet{Maglio2004} (orange circle), \citet{gilli05} (blue triangle), \citet{gilli} (purple square), \citet{maglio2017} (green diamonds), \citet{hale_cosmos} (cyan circles) \& \citet{arnab2020} (magenta inverted triangle) have been plotted. Predictions from SKADS \citep{skads} have also been plotted for Radio Quiet Quasars (RQQ), FR-I and FR-II radio galaxies. \textbf{(right panel)} Spatial correlation length ($\mathrm{r_{0}}$) SFGs as a function of redshift (red circle). For comparison, values obtained by \citet{gilli07} (black circles), \citet{Starikova_2012}(maroon diamonds), \citet{maglio2013} (green inverted triangles), \citet{Dolley_2014} (blue diamonds), \citet{maglio2017} (green circles), \citet{hale_cosmos} (cyan pentagon), \& \citet{arnab2020} (magenta triangles) have been plotted. Predictions from SKADS \citep{skads} have also been plotted for SFGs and Starburst galaxies (SB).}
\label{r0_sep}
\end{figure*}

\begin{figure}
\includegraphics[width=\columnwidth,height=3.0in]{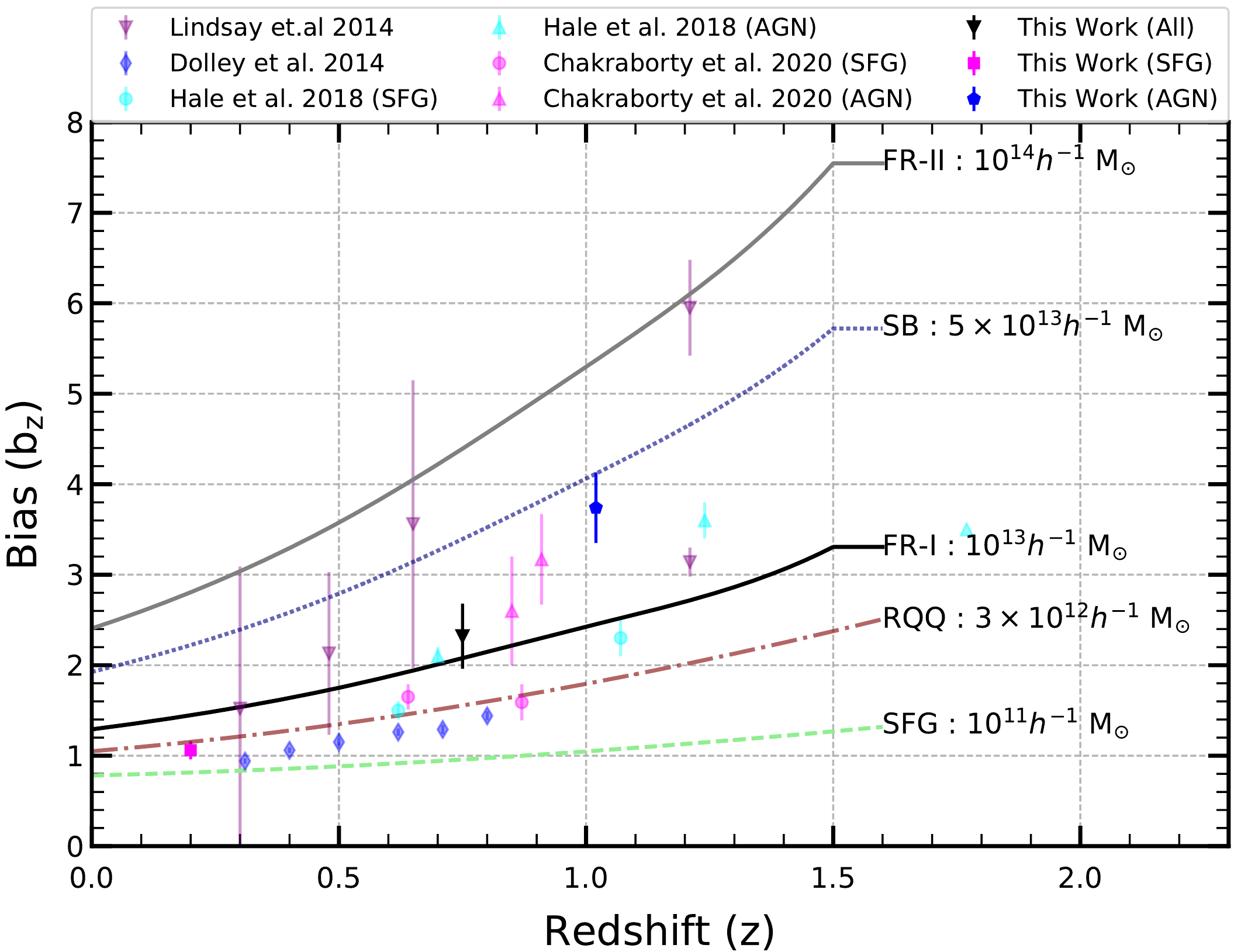}
 \caption{Bias parameter $b_z$ for whole population (black inverted triangle) as well AGNs (blue pentagons) and SFGs (magenta squares). Bias parameters from previous observations by \citet{lindsay_first} (purple inverted traingles), \citet{Dolley_2014} (blue diamonds), \citet{hale_cosmos} (cyan triangles for AGNs and cyan circles SFGs) \& \citet{arnab2020} (magenta triangles and circles for AGNs and SFGs respectively) are shown.  Predictions from SKADS is also shown by the continuous curves.}
\label{bias}
\end{figure}

Figure \ref{r0_sep} shows the observationally determined values from surveys at various wavebands for $\mathrm{r_{0}}$ as a function of their redshift, while Figure \ref{bias} shows the same for the bias parameter. The left and right panels of Figure \ref{r0_sep} are for AGNs and SFGs, respectively. Table \ref{spatial_table} summarizes the values obtained in radio surveys only, while Figures \ref{r0_sep}, \ref{bias} show the observed values for surveys at radio as well as other wavebands, e.g. IR and X-Ray. 

The clustering length for AGNs in this work is  at z$_{median} \approx$1.02 is $\textrm 8.30^{+0.96}_{-0.91}$ Mpc $\rm h^{-1}$. Using X-Ray selected AGNs in the COSMOS field, \citealt{gilli} obtained clustering lengths at redshifts upto $\sim$3.0. They divided their sample into a number of bins, to obtain $\mathrm{r_{0}}$ at different median redshifts. For their entire sample, taking slope of the angular correlation function as 1.80, $\mathrm{r_{0}}$ was $\textrm 8.39^{+0.41}_{-0.39}$ Mpc $\rm h^{-1}$, for a median redshift of 0.98. It is consistent with the value obtained here at a similar redshift. The clustering length with this work is also consistent within error bars for AGNs at 400 MHz and 610 MHz of \citet{arnab2020}.
For their work, they obtain an $\mathrm{r_{0}}$ value of $\textrm 7.30^{+1.14}_{-1.12}$ Mpc $\rm h^{-1}$ at $z \approx0.91$, $\textrm 6.00^{+1.5}_{-1.3}$ Mpc $\rm h^{-1}$ at $z_{median} = 0.84$. The clustering length estimates for radio selected AGNs in the COSMOS field at 1.4 GHz \citep{maglio2017} and 3 GHz  \citep{hale_cosmos} observed with the VLA also agree within error bars with the estimates obtained here. \citet{maglio2017} have found a clustering length of $\textrm 7.84^{+1.75}_{-2.31}$ Mpc $\rm h^{-1}$ at z $\approx$1.25 while \citep{hale_cosmos} obtained $\textrm 6.90^{+0.60}_{-0.70}$ Mpc $\rm h^{-1}$, $\textrm 9.60^{+0.70}_{-0.70}$ Mpc $\rm h^{-1}$ and $\textrm 7.30^{+0.90}_{-0.90}$ Mpc $\rm h^{-1}$ at $z\approx$ 0.70, 1.24, 1.77 respectively. Using X-ray selected AGNs in the CDFS field, \citet{gilli05} obtained a value of $\textrm 10.30^{+1.7}_{-1.7}$ Mpc $\rm h^{-1}$ at z $\approx$ 0.84. This value though higher than the values for radio selected AGNs, is still consistent within error bars.

For the SFGs population (right panel of Figure \ref{r0_sep}), the median redshift is 0.20. At this redshift, the clustering length is  $\textrm 3.22^{+0.34}_{-0.32}$ Mpc $\rm h^{-1}$. This estimate is at a redshift lower than previous observations. An extensive study at mid-IR frequency has been done by \citet{Dolley_2014} for SFGs. The lowest redshift probed in their study is 0.31, where $\mathrm{r_{0}}$ is  $\textrm 3.41^{+0.18}_{-0.18}$ Mpc $\rm h^{-1}$. Thus, the value is consistent with that obtained here at a nearby redshift. \citet{maglio2013} studied the clustering of SFGs using the Herschel PACS Evolutionary Probe observations of the COSMOS and Extended Groth Strip fields. They found clustering lengths for SFGs out to $z \approx$ 2. For the ELAIS-N1 field at 400 MHz and 610 MHz, \citet{arnab2020} reported clustering length of $\textrm 4.62^{+0.39}_{-0.40}$ Mpc $\rm h^{-1}$ and $\textrm 4.16^{+0.70}_{-0.80}$ Mpc $\rm h^{-1}$ at redshifts 0.64 and 0.87 respectively. The 3 GHz COSMOS field studies of \citet{hale_cosmos} gave clustering lengths $\textrm 5.00^{+0.50}_{-0.60}$ Mpc $\rm h^{-1}$ and  $\textrm 6.1^{+0.60}_{-0.70}$ Mpc $\rm h^{-1}$ respectively at z$\approx$ 0.62 and 1.07. The mid-IR selected samples for Lockman Hole give $\mathrm{r_{0}}$ values $\textrm 4.98^{+0.28}_{-0.28}$ Mpc and $\textrm 8.04^{+0.69}_{-0.69}$ Mpc $\rm h^{-1}$ at $z \approx$ 0.7 and 1.7 respectively. Similarly, the mid-IR sample for \citet{gilli07} has clustering lengths $\mathrm{r_{0}}$ is $\textrm 4.25^{+0.12}_{-0.12}$ Mpc $\rm h^{-1}$ and $\textrm 3.81^{+0.10}_{-0.10}$ Mpc $\rm h^{-1}$ for $z \approx$ 0.67 and 0.73.



The results have also been compared with the assumed bias models of the semi-empirical simulated catalogue of the extragalactic sky, the Square Kilometer Array Design Studies (referred to as SKADS henceforth, \citealt{skads}). This simulation models the large-scale cosmological distribution of radio sources to aid the design of next-generation radio interferometers. It covers a sky area of $20^\circ \times 20 ^\circ$, with sources out to a cosmological redshift of $z\sim$20 and a minimum flux 10\,nJy at 151, 610 MHz \& 1.4, 4.86 and 18 GHz. The simulated sources are drawn from observed and, in some cases, extrapolated luminosity functions on an underlying dark matter density field with biases to reflect the measured large-scale clustering. It uses a numerical Press–Schechter \citep{PressSchechter1974} style filtering on the density field to identify clusters of galaxies. The SKADS catalogue has been used here for statistical inference of the spatial and angular clustering variations of the sources with redshift. It should be mentioned here that the T-RECS catalogue \citep{trecs} incorporates more updated results from the recent observations. However, the evolution of the bias parameter and clustering length with redshift is not available for the same; hence SKADS has been used.


The bias parameter for AGNs and SFGs at z$_{median}$ 1.02 and 0.20 is $3.74^{+0.39}_{-0.36}$ and $1.06^{+0.10}_{-0.10}$ respectively. Although the value for AGNs is slightly higher than those obtained by \citet{hale_cosmos} and \citet{arnab2020}, it is still with reasonable agreement with the SKADS for FR-I galaxies. Comparison with the population distribution of the SKADS simulation of \citet{skads}, in terms of both clustering length and the bias parameter (solid magenta pentagons in Figure \ref{bias}) show that the AGN population is dominated by FR-I type galaxies hosted in massive haloes with $\sim M_{h}$= 5$\times$10$^{13}\rm h^{-1}M_{\odot}$. It can also seen from Figure \ref{bias}, that the mass of the haloes hosting the SFG samples of the current sample is $\sim M_{h}$= 3$\times10^{12} \rm h^{-1}M_{\odot}$. Thus, it is seen that the SFGs have a lower range of halo masses compared to AGNs, which implies that the latter inhabits more massive haloes and are more biased tracers of the dark matter density field.

\section{Discussion}
\label{discussion}

The analysis of clustering properties of radio selected sources in the Lockman Hole region presented in this work is one of the first results reported at 325 MHz. Similar studies were previously done at 400 MHz for the ELAIS-N1 field \citep{arnab2020}, however for a much smaller area. Beside analysing at a mostly unexplored frequency, this work also presents a comparatively large area with a significant number of sources. Previous clustering study of the same area using 1.4 GHz data from WSRT by \citet{lh_clustering_1.4} used 1173 sources with a flux density cut-off of 0.12 mJy at 1.4 GHz (or 0.4 mJy at 325 MHz). Their obtained clustering amplitude is slightly higher than previous surveys, as acknowledged by the authors. However, further investigation is required to ascertain the reason for the deviation. Clustering analysis was also done using the recent LoTSS observation of the HETDEX spring field \citep{lotss_clustering}. The clustering analyses were produced with many flux density cut-offs and masks, and the most reliable estimate was for a flux density limit of 2 mJy at 150 MHz (or $\sim$1.0 mJy at 325 MHz). The clustering amplitude estimate at 2 mJy limit for LoTSS is consistent with that obtained here within error bars. As seen from Figure \ref{angular_all}, the clustering amplitude obtained here agrees with previous observations. The slightly higher values for the bias parameter of the AGNs for this work, compared to that of \citet{hale_cosmos} and \citet{arnab2020}, may be attributed to the different flux limits of the studies. This implies that each of these observations are probing slightly different populations of sources (with slightly different luminosities as discussed later). Nevertheless, as seen from Figure \ref{bias}, the values are broadly consistent with each other.

The angular clustering amplitude for this work as shown in Figure \ref{angular_sep} agree with previous observations, as seen in Table \ref{spatial_table}. The angular clustering of sources for this work are calculated with \texttt{TreeCorr} using the default values for most parameters. It has been shown in \citet{lotss_clustering} that using default value of 1 for the parameter \texttt{bin\_slop} gives less accurate results than for values $\leq$1. \citet{lotss_clustering} obtained the most accurate values for \texttt{bin\_slop}=0. They also showed that angular clustering amplitudes deviate largely from precise values (calculated from a separate brute-force algorithm, see \citep{lotss_clustering} for details) at angular scales $\gtrsim$1$^\circ$. However, the computation times also significantly increased for \texttt{bin\_slop}=0. The use of default parameters in this work might be the cause slight oscillation of the correlation function seen around the best fit curves. Nevertheless, since results obtained here are in reasonable agreement with the previous observations and owing to constraints in the available computing power, the default values have been used.

The clustering lengths and bias parameters obtained here also agree with previous studies, as evident from Table \ref{spatial_table} and Figures \ref{r0_sep}, \ref{bias}. Comparison of bias parameter with SKADS simulation \citealt{skads} shows that the expected mass for dark matter haloes hosting AGNs is orders of magnitude higher than that for SFGs. This trend is consistent with previous observations (see for example \citealt{maglio2017, hale_cosmos, arnab2020}). Studies on the luminosity of the AGNs and SFGs suggest that it is correlated with source clustering, and hence with the bias parameter and host halo mass \citep{hale_cosmos}. Using observations of high and low luminosity AGNs in the COSMOS field, \citet{hale_cosmos} showed that the luminosity and clustering are correlated, with the higher luminosity AGNs residing in more massive galaxies \citep{jarvis2001a}. Thus, they are hosted by more massive haloes. Again, this points to AGNs (which are in general more luminous than SFGs) being hosted by more massive haloes. However, it should also be mentioned that the two population of AGNs in \citet{hale_cosmos} studied are at different redshifts, with the low excitation population studied at $z\lesssim0.65$. So it is possible that this population may evolve into higher mass haloes at higher redshifts. Moreover, some works (for example \citealt{Mendez_2016}) do not find any relation between clustering and luminosity. Nevertheless, following \citet{hale_cosmos}, studies using a larger population of samples covering a large range of luminosities is required for probing the relationship with clustering. 



It is also observed in Figure \ref{bias} that the bias parameter for SFGs (solid blue square) is higher than the SKADS predicted values for this population. This trend is consistent with that observed in previous studies of \citet{hale_cosmos} (cyan circles) and \citet{arnab2020} (light magenta triangles). The trend observed in \citet{Dolley_2014} (light blue diamonds) is almost similar as well. There may be two reasons that cause this variation- contamination of the SFG sample by star-burst (SB) galaxies or underestimation of halo mass for SKADS. If there exists a few SB samples in the SFG population, comparison with SKADS (blue dotted curve in Figure \ref{bias}) shows that the overall value for bias (as well as $\mathrm{r_{0}}$) will be higher than an uncontaminated sample. However, the most likely reason remains the second one, the halo mass used in the SKADS is not a correct representation, which has also been hinted at by comparing the values obtained from previous observations (for instance \citet{Dolley_2014, hale_cosmos, arnab2020}) in  Figure \ref{bias}. However, it is also seen from Figure \ref{r0_sep} spatial clustering length for SFGs agrees with SKADS. The exact reason for agreement of $\mathrm{r_{0}}$ and disagreement of $b(z)$ for SFGs in this work with SKADS is unclear, and will be investigated in detail in later works.

Analyses like the one presented here are important for fully understanding how bias scales with redshift as well as with source properties like luminosity. This is important for cosmology, since the bias relates to dark matter distribution, and is thus essential for understanding the underlying cosmological parameters that define the Universe. 

It should be mentioned here that the current study also has certain limitations. Due to unknown systematics at large scales, and lack of optical matches, the entire observed field could not be utilised for this study. Additionally, the source classification is done based solely on the radio luminosity of the sources. As has already been mentioned previously, \citet{maglio2014} showed that the chances of contamination between the populations is very less using this criteria. Nonetheless, there are several other methods that can also be used to classify AGNs and SFGs in a sample (detailed in Section \ref{classify}). Future works will present a more detailed analysis using the different multi-wavelength classification schemes available.Such multi-frequency studies, combined with the present work and similar studies with other fields will enhance the knowledge of the extragalactic sources and provide more insights into the processes governing their formation and evolution.

\section{Conclusion}
\label{conclusion}

This work investigates the higher-order source statistics, namely the angular and spatial clustering of the sources detected in the Lockman Hole field. The data was observed by the legacy GMRT at 325 MHz. The details of data analysis and catalogue extraction are discussed in \citet{aishrila1}. The initial step involved merging the multi-component sources present in the raw catalogue. The resultant catalogue was cross-matched with SDSS and HELP catalogues to identify sources with either spectroscopic or photometric redshift information. A region of radius 1.8$^\circ$ around the phase center was selected for optical identifications, yielding $\sim$95\% matches. All the sources with redshift distribution were separated into AGN and SFG populations using the criterion for radio luminosity. The angular correlation function was determined for the combined population for separation between $36\arcsec$ to 2$^\circ$. A power law fitted to this function, keeping a fixed power law index of $\mathrm{\gamma = 1.80}$, as estimated theoretically \citep{Peebles1980}. This gave the value  of clustering amplitude $\mathrm{log_{10}(A)} = -2.73^{0.11}_{-0.15}$. 


The source population was further divided into AGNs and SFGs based on their radio luminosity, and clustering analyses were done for these populations as well. Using the redshift information and the clustering amplitude, spatial correlation length was determined using Limber inversion for the AGNs and SFGs.  The correlation length and bias parameters have been obtained for the full sample, as well as the classified AGN and SFG population. For the full sample at z$_{median}\approx$ 0.78, $\mathrm{r_{0}}$ = 3.50$\mathrm{^{+0.50}_{-0.50}}$ Mpc h$^\textrm{-1}$ and $b(z)$ = 2.22$\mathrm{^{+0.33}_{-0.36}}$. For AGNs,  the values are $\mathrm{r_{0}}$ = 8.30$\mathrm{^{+0.96}_{-0.91}}$ Mpc h$\rm ^{-1}$ and $b(z)$ = 3.74$\mathrm{^{+0.39}_{-0.36}}$ at z$_{median}\approx$ 1.02. At z$_{median}\approx$ 0.20, SFGs have values $\mathrm{r_{0}}$ = 3.22$\mathrm{^{+0.34}_{-0.32}}$ Mpc h$\rm ^{-1}$ and $b(z)$ = 1.06$\mathrm{^{+0.10}_{-0.10}}$. The clustering length for AGNs is reasonably consistent with \citet{gilli} as well as \citet{arnab2020}. For SFGs, the values are consistent with \citet{Dolley_2014}.  


The obtained values have also been compared with the SKADS simulation of \citet{skads}. The comparative analysis suggests that the AGNs are dominated by FR-I galaxies, with host dark matter halo masses of $M_{h}$=5-6 $\times$ 10$^{13}h^{-1}M_{\odot}$. For SFGs, the estimated halo mass obtained from SKADS is lower compared to the value $\sim M_{h}$=3 $\times10^{12}h^{-1}M_{\odot}$ obtained here. The halo mass obtained here are in agreement with previous literature \citep{Dolley_2014, hale_cosmos, arnab2020}. It is worthwhile to mention that while the current classifications are based on radio luminosity for each source alone, the results are in agreement with clustering properties for populations of AGNs and SFGs in X-Ray and mid-IR surveys as well. However, there are some deviations from the predictions of the SKADS simulation. This deviation, seen in other observations as well, emphasizes the need for wider and deeper low-frequency observations. These will be able to constrain the host properties better, leading to better models formation and evolution of the different sources, and better understanding of the distribution of these populations over space and time (redshift). 




The study done in this work aims to characterize the clustering and bias of an observed population of radio-selected sources, both as a combined population and as distinct classes of sources (namely AGNs and SFGs). This work is the first to report the clustering properties of radio-selected sources at 325 MHz. Thus, current data, being at a frequency with little previous clustering study, bridges the gap between low frequency and high-frequency studies. It also has the advantage of covering a wider area than many recent studies with moderately deep RMS values, thus probing a larger number of sources with fluxes at the sub-mJy level ($\sim$0.3 mJy). More such studies using observational data are required for constraining cosmology and probing how different source populations are influenced by their parent halos and how they evolve with time (redshift).
Additionally, for sensitive observations of CD/EoR and post-EoR science, accurate and realistic models for compact source populations that comprise a significant fraction of foregrounds are required. Studies on the effect of imperfections in foreground modeling in the power spectrum estimates will require detailed observational studies of source position and flux distributions. For realistic estimates, second-order statistics like clustering also cannot be ignored. Thus many more analyses like the one done in this paper will be required for better understanding the effects of the interplay between the various cosmological parameters on the different populations of sources and putting constraints on said parameters. Sensitive large area surveys like the MIGHTEE \citep{Jarvis2016} with the MeerKat telescope, LoTSS \citep{lotss_dr1, tasse2020, sabater} with the LOFAR, EMU \citep{Norris2011, norris2021} with the ASKAP, as well as those to be done with the upcoming SKA-mid telescope would provide wider as well as deeper data for doing cosmology. The current work demonstrates that even instruments like the legacy GMRT can provide reasonable data depth and coverage for
cosmological observations.

\section*{ACKNOWLEDGEMENTS}
AM would like to thank Indian Institute of Technology Indore for supporting this research with Teaching Assistantship. AM further acknowledges Akriti Sinha for helpful suggestions and pointing to the HELP catalogue and Sumanjit Chakraborty for helpful discussions. The authors thank the staff of GMRT for making this observation possible. GMRT is run by National Centre for Radio Astrophysics of the Tata Institute of Fundamental Research. The authors also acknowledge SDSS for the spectroscopic redshift catalogues. Funding for SDSS-III has been provided by the Alfred P. Sloan Foundation, the Participating Institutions, the National Science Foundation, and the U.S. Department of Energy Office of Science. The SDSS-III web site is \url{http://www.sdss3.org/}. The authors thank the anonymous reviewer for their thorough review that has helped to improve the quality of the work.

\section*{Data Availability}
The raw data for this study is available in the GMRT archive (\url{https://naps.ncra.tifr.res.in/goa/data/search}). The spectroscopic redshift data is available from \url{https://skyserver.sdss.org/casjobs/}, and the photometric redshift data is available from \url{https://hedam.lam.fr/HELP/}.The P{\tiny Y}BDSF catalogue used here (accompanying \citet{aishrila1}) is available on VizieR at \url{https://cdsarc.cds.unistra.fr/viz-bin/cat/J/MNRAS/495/4071}.

\section*{Software}

This work relies on the Python programming language (\url{https://www.python.org/}). The packages used here are astropy (\url{https://www.astropy.org/}; \citet{astropy:2013,astropy:2018}), numpy (\url{https://numpy.org/}), scipy (\url{https://www.scipy.org/}), matplotlib (\url{https://matplotlib.org/}), TreeCorr (\url{https://github.com/rmjarvis/TreeCorr}).



\bibliographystyle{mnras}
\bibliography{references} 

\begin{thebibliography}{}
\makeatletter
\relax
\def\mn@urlcharsother{\let\do\@makeother \do\$\do\&\do\#\do\^\do\_\do\%\do\~}
\def\mn@doi{\begingroup\mn@urlcharsother \@ifnextchar [ {\mn@doi@}
  {\mn@doi@[]}}
\def\mn@doi@[#1]#2{\def\@tempa{#1}\ifx\@tempa\@empty \href
  {http://dx.doi.org/#2} {doi:#2}\else \href {http://dx.doi.org/#2} {#1}\fi
  \endgroup}
\def\mn@eprint#1#2{\mn@eprint@#1:#2::\@nil}
\def\mn@eprint@arXiv#1{\href {http://arxiv.org/abs/#1} {{\tt arXiv:#1}}}
\def\mn@eprint@dblp#1{\href {http://dblp.uni-trier.de/rec/bibtex/#1.xml}
  {dblp:#1}}
\def\mn@eprint@#1:#2:#3:#4\@nil{\def\@tempa {#1}\def\@tempb {#2}\def\@tempc
  {#3}\ifx \@tempc \@empty \let \@tempc \@tempb \let \@tempb \@tempa \fi \ifx
  \@tempb \@empty \def\@tempb {arXiv}\fi \@ifundefined
  {mn@eprint@\@tempb}{\@tempb:\@tempc}{\expandafter \expandafter \csname
  mn@eprint@\@tempb\endcsname \expandafter{\@tempc}}}

\bibitem[\protect\citeauthoryear{Abdalla et~al.,}{Abdalla
  et~al.}{2015}]{abdalla2015cosmology}
Abdalla F.~B.,  et~al., 2015, Cosmology from HI galaxy surveys with the SKA
  (\mn@eprint {arXiv} {1501.04035})

\bibitem[\protect\citeauthoryear{Afonso, Georgakakis, Almeida, Hopkins, Cram,
  Mobasher  \& Sullivan}{Afonso et~al.}{2005}]{Afonso_2005}
Afonso J.,  Georgakakis A.,  Almeida C.,  Hopkins A.~M.,  Cram L.~E.,  Mobasher
  B.,   Sullivan M.,  2005, \mn@doi [The Astrophysical Journal]
  {10.1086/428923}, 624, 135

\bibitem[\protect\citeauthoryear{Alam et~al.,}{Alam et~al.}{2021}]{sdss_hz}
Alam S.,  et~al., 2021, \mn@doi [Phys. Rev. D] {10.1103/PhysRevD.103.083533},
  103, 083533

\bibitem[\protect\citeauthoryear{Ali, Bharadwaj  \& Chengalur}{Ali
  et~al.}{2008}]{ali08}
Ali S.~S.,  Bharadwaj S.,   Chengalur J.~N.,  2008, \mn@doi [MNRAS]
  {10.1111/j.1365-2966.2008.12984.x}, 385, 2166

\bibitem[\protect\citeauthoryear{Allison et~al.,}{Allison
  et~al.}{2015}]{allison}
Allison R.,  et~al., 2015, \mn@doi [Monthly Notices of the Royal Astronomical
  Society] {10.1093/mnras/stv991}, 451, 849

\bibitem[\protect\citeauthoryear{{Astropy Collaboration} et~al.,}{{Astropy
  Collaboration} et~al.}{2013}]{astropy:2013}
{Astropy Collaboration} et~al., 2013, \mn@doi [\aap]
  {10.1051/0004-6361/201322068}, \href
  {http://adsabs.harvard.edu/abs/2013A%26A...558A..33A} {558, A33}

\bibitem[\protect\citeauthoryear{Bardeen, Bond, Kaiser  \& Szalay}{Bardeen
  et~al.}{1986}]{Bardeen1986}
Bardeen J.~M.,  Bond J.~R.,  Kaiser N.,   Szalay A.~S.,  1986, \mn@doi [The
  Astrophysical Journal] {10.1086/164143}, 304, 15

\bibitem[\protect\citeauthoryear{Becker, White  \& Helfand}{Becker
  et~al.}{1995}]{FIRST}
Becker R.~H.,  White R.~L.,   Helfand D.~J.,  1995, \mn@doi [ApJ]
  {10.1086/176166}, 450, 559

\bibitem[\protect\citeauthoryear{Bell}{Bell}{2003}]{Bell2003}
Bell E.~F.,  2003, \mn@doi [The Astrophysical Journal] {10.1086/367829}, 586,
  794

\bibitem[\protect\citeauthoryear{Best \& Heckman}{Best \& Heckman}{2012}]{best}
Best P.~N.,  Heckman T.~M.,  2012, \mn@doi [Monthly Notices of the Royal
  Astronomical Society] {10.1111/j.1365-2966.2012.20414.x}, 421, 1569

\bibitem[\protect\citeauthoryear{Blake \& Wall}{Blake \&
  Wall}{2002a}]{blake_wall_2002a}
Blake C.,  Wall J.,  2002a, \mn@doi [Monthly Notices of the Royal Astronomical
  Society] {10.1046/j.1365-8711.2002.05163.x}, 329, L37

\bibitem[\protect\citeauthoryear{Blake \& Wall}{Blake \&
  Wall}{2002b}]{blake_wall}
Blake C.,  Wall J.,  2002b, \mn@doi [Monthly Notices of the Royal Astronomical
  Society] {10.1046/j.1365-8711.2002.05979.x}, 337, 993

\bibitem[\protect\citeauthoryear{Blake, Abdalla, Bridle  \& Rawlings}{Blake
  et~al.}{2004}]{BLAKE20041063}
Blake C.,  Abdalla F.,  Bridle S.,   Rawlings S.,  2004, \mn@doi [New Astronomy
  Reviews] {https://doi.org/10.1016/j.newar.2004.09.045}, 48, 1063

\bibitem[\protect\citeauthoryear{Blanton et~al.,}{Blanton
  et~al.}{2017}]{Blanton2017}
Blanton M.~R.,  et~al., 2017, \mn@doi [The Astronomical Journal]
  {10.3847/1538-3881/aa7567}, 154, 28

\bibitem[\protect\citeauthoryear{Bonaldi, Bonato, Galluzzi, Harrison, Massardi,
  Kay, De Zotti  \& Brown}{Bonaldi et~al.}{2018}]{trecs}
Bonaldi A.,  Bonato M.,  Galluzzi V.,  Harrison I.,  Massardi M.,  Kay S.,
  De Zotti G.,   Brown M.~L.,  2018, \mn@doi [Monthly Notices of the Royal
  Astronomical Society] {10.1093/mnras/sty2603}, 482, 2

\bibitem[\protect\citeauthoryear{Bonato, Prandoni, De Zotti, Brienza, Morganti
   \& Vaccari}{Bonato et~al.}{2020}]{lh_clustering_1.4}
Bonato M.,  Prandoni I.,  De Zotti G.,  Brienza M.,  Morganti R.,   Vaccari
  M.,  2020, \mn@doi [Monthly Notices of the Royal Astronomical Society]
  {10.1093/mnras/staa3218}, 500, 22

\bibitem[\protect\citeauthoryear{Bonato et~al.,}{Bonato
  et~al.}{2021}]{bonato2021lofar}
Bonato M.,  et~al., 2021, The LOFAR Two-metre Sky Survey Deep fields: A new
  analysis of low-frequency radio luminosity as a star-formation tracer in the
  Lockman Hole region (\mn@eprint {arXiv} {2109.06735})

\bibitem[\protect\citeauthoryear{Bonzini, Padovani, Mainieri, Kellermann,
  Miller, Rosati, Tozzi  \& Vattakunnel}{Bonzini et~al.}{2013}]{bonzini}
Bonzini M.,  Padovani P.,  Mainieri V.,  Kellermann K.~I.,  Miller N.,  Rosati
  P.,  Tozzi P.,   Vattakunnel S.,  2013, \mn@doi [Monthly Notices of the Royal
  Astronomical Society] {10.1093/mnras/stt1879}, 436, 3759

\bibitem[\protect\citeauthoryear{Calistro~Rivera et~al.,}{Calistro~Rivera
  et~al.}{2017}]{rivera}
Calistro~Rivera G.,  et~al., 2017, \mn@doi [Monthly Notices of the Royal
  Astronomical Society] {10.1093/mnras/stx1040}, 469, 3468

\bibitem[\protect\citeauthoryear{Camacho et~al.,}{Camacho et~al.}{2019}]{des}
Camacho H.,  et~al., 2019, \mn@doi [Monthly Notices of the Royal Astronomical
  Society] {10.1093/mnras/stz1514}, 487, 3870

\bibitem[\protect\citeauthoryear{Camera, Santos, Bacon, Jarvis, McAlpine,
  Norris, Raccanelli  \& Röttgering}{Camera et~al.}{2012}]{camera}
Camera S.,  Santos M.~G.,  Bacon D.~J.,  Jarvis M.~J.,  McAlpine K.,  Norris
  R.~P.,  Raccanelli A.,   Röttgering H.,  2012, \mn@doi [Monthly Notices of
  the Royal Astronomical Society] {10.1111/j.1365-2966.2012.22073.x}, 427, 2079

\bibitem[\protect\citeauthoryear{Carilli \& Rawlings}{Carilli \&
  Rawlings}{2004}]{CARILLI2004979}
Carilli C.,  Rawlings S.,  2004, \mn@doi [New Astronomy Reviews]
  {https://doi.org/10.1016/j.newar.2004.09.001}, 48, 979

\bibitem[\protect\citeauthoryear{Carvalho, Bernui, Benetti, Carvalho  \&
  Alcaniz}{Carvalho et~al.}{2016}]{sdss_10_bao}
Carvalho G.~C.,  Bernui A.,  Benetti M.,  Carvalho J.~C.,   Alcaniz J.~S.,
  2016, \mn@doi [Phys. Rev. D] {10.1103/PhysRevD.93.023530}, 93, 023530

\bibitem[\protect\citeauthoryear{Chakraborty, Dutta, Datta  \& Roy}{Chakraborty
  et~al.}{2020}]{arnab2020}
Chakraborty A.,  Dutta P.,  Datta A.,   Roy N.,  2020, \mn@doi [Monthly Notices
  of the Royal Astronomical Society] {10.1093/mnras/staa945}, 494, 3392

\bibitem[\protect\citeauthoryear{Chambers et~al.,}{Chambers
  et~al.}{2019}]{panstarrs1}
Chambers K.~C.,  et~al., 2019, The Pan-STARRS1 Surveys (\mn@eprint {arXiv}
  {1612.05560})

\bibitem[\protect\citeauthoryear{Condon}{Condon}{1989}]{Condon1989}
Condon J.~J.,  1989, \mn@doi [The Astrophysical Journal] {10.1086/167176}, 338,
  13

\bibitem[\protect\citeauthoryear{Condon, Cotton, Greisen, Yin, Perley, Taylor
  \& Broderick}{Condon et~al.}{1998}]{Condon1998}
Condon J.~J.,  Cotton W.~D.,  Greisen E.~W.,  Yin Q.~F.,  Perley R.~A.,  Taylor
  G.~B.,   Broderick J.~J.,  1998, \mn@doi [The Astronomical Journal]
  {10.1086/300337}, 115, 1693

\bibitem[\protect\citeauthoryear{Cooray \& Furlanetto}{Cooray \&
  Furlanetto}{2004}]{Cooray2004}
Cooray A.,  Furlanetto S.~R.,  2004, \mn@doi [ApJ] {10.1086/421241}, 606, L5

\bibitem[\protect\citeauthoryear{Cress, Helfand, Becker, Gregg  \& White}{Cress
  et~al.}{1996}]{Cress_1996}
Cress C.~M.,  Helfand D.~J.,  Becker R.~H.,  Gregg M.~D.,   White R.~L.,  1996,
  \mn@doi [The Astrophysical Journal] {10.1086/178122}, 473, 7

\bibitem[\protect\citeauthoryear{Cucciati et~al.,}{Cucciati
  et~al.}{2012}]{cucciati}
Cucciati O.,  et~al., 2012, \mn@doi [A\&A] {10.1051/0004-6361/201118010}, 539,
  A31

\bibitem[\protect\citeauthoryear{Davies et~al.,}{Davies
  et~al.}{2016}]{Davies2017}
Davies L. J.~M.,  et~al., 2016, \mn@doi [Monthly Notices of the Royal
  Astronomical Society] {10.1093/mnras/stw3080}, 466, 2312

\bibitem[\protect\citeauthoryear{DeBoer et~al.,}{DeBoer et~al.}{2009}]{askap}
DeBoer D.~R.,  et~al., 2009, \mn@doi [Proceedings of the IEEE]
  {10.1109/JPROC.2009.2016516}, 97, 1507

\bibitem[\protect\citeauthoryear{Delhaize et~al.,}{Delhaize
  et~al.}{2017}]{Delhaize2017}
Delhaize J.,  et~al., 2017, \mn@doi [A\&A] {10.1051/0004-6361/201629430}, 602,
  A4

\bibitem[\protect\citeauthoryear{Delvecchio et~al.,}{Delvecchio
  et~al.}{2021}]{Delvecchio2021}
Delvecchio I.,  et~al., 2021, \mn@doi [A\&A] {10.1051/0004-6361/202039647},
  647, A123

\bibitem[\protect\citeauthoryear{Desjacques, Jeong  \& Schmidt}{Desjacques
  et~al.}{2018}]{DESJACQUES20181}
Desjacques V.,  Jeong D.,   Schmidt F.,  2018, \mn@doi [Physics Reports]
  {https://doi.org/10.1016/j.physrep.2017.12.002}, 733, 1

\bibitem[\protect\citeauthoryear{Di~Matteo, Ciardi  \& Miniati}{Di~Matteo
  et~al.}{2004}]{dimatteo2004}
Di~Matteo T.,  Ciardi B.,   Miniati F.,  2004, \mn@doi [MNRAS]
  {10.1111/j.1365-2966.2004.08443.x}, 355, 1053

\bibitem[\protect\citeauthoryear{Dolley et~al.,}{Dolley
  et~al.}{2014}]{Dolley_2014}
Dolley T.,  et~al., 2014, \mn@doi [The Astrophysical Journal]
  {10.1088/0004-637x/797/2/125}, 797, 125

\bibitem[\protect\citeauthoryear{Donley et~al.,}{Donley
  et~al.}{2012}]{Donley_2012}
Donley J.~L.,  et~al., 2012, \mn@doi [The Astrophysical Journal]
  {10.1088/0004-637x/748/2/142}, 748, 142

\bibitem[\protect\citeauthoryear{Donoso, Yan, Stern  \& Assef}{Donoso
  et~al.}{2014}]{Donoso2014}
Donoso E.,  Yan L.,  Stern D.,   Assef R.~J.,  2014, \mn@doi [The Astrophysical
  Journal] {10.1088/0004-637X/789/1/44}, 789, 44

\bibitem[\protect\citeauthoryear{Duncan et~al.,}{Duncan et~al.}{2017}]{help2}
Duncan K.~J.,  et~al., 2017, \mn@doi [Monthly Notices of the Royal Astronomical
  Society] {10.1093/mnras/stx2536}, 473, 2655

\bibitem[\protect\citeauthoryear{Duncan, Jarvis, Brown  \& Röttgering}{Duncan
  et~al.}{2018}]{duncan2018}
Duncan K.~J.,  Jarvis M.~J.,  Brown M. J.~I.,   Röttgering H. J.~A.,  2018,
  \mn@doi [Monthly Notices of the Royal Astronomical Society]
  {10.1093/mnras/sty940}, 477, 5177

\bibitem[\protect\citeauthoryear{Dunlop \& Peacock}{Dunlop \&
  Peacock}{1990}]{Dunlop1990}
Dunlop J.~S.,  Peacock J.~A.,  1990, Monthly Notices of the Royal Astronomical
  Society, 247, 19

\bibitem[\protect\citeauthoryear{Dye et~al.,}{Dye et~al.}{2017}]{uhs}
Dye S.,  et~al., 2017, \mn@doi [Monthly Notices of the Royal Astronomical
  Society] {10.1093/mnras/stx2622}, 473, 5113

\bibitem[\protect\citeauthoryear{Eddington}{Eddington}{1913}]{Eddington}
Eddington A.~S.,  1913, \mn@doi [MNRAS] {10.1093/mnras/73.5.359}, 73, 359

\bibitem[\protect\citeauthoryear{Eisenstein, Hu  \& Tegmark}{Eisenstein
  et~al.}{1999}]{Eisenstein_1999}
Eisenstein D.~J.,  Hu W.,   Tegmark M.,  1999, \mn@doi [The Astrophysical
  Journal] {10.1086/307261}, 518, 2

\bibitem[\protect\citeauthoryear{Eisenstein et~al.,}{Eisenstein
  et~al.}{2005}]{einstein_bao}
Eisenstein D.~J.,  et~al., 2005, \mn@doi [The Astrophysical Journal]
  {10.1086/466512}, 633, 560

\bibitem[\protect\citeauthoryear{Eisenstein et~al.,}{Eisenstein
  et~al.}{2011}]{SDSSIII}
Eisenstein D.~J.,  et~al., 2011, \mn@doi [The Astronomical Journal]
  {10.1088/0004-6256/142/3/72}, 142, 72

\bibitem[\protect\citeauthoryear{Fanaroff \& Riley}{Fanaroff \&
  Riley}{1974}]{fanaroff_riley}
Fanaroff B.~L.,  Riley J.~M.,  1974, \mn@doi [Monthly Notices of the Royal
  Astronomical Society] {10.1093/mnras/167.1.31P}, 167, 31P

\bibitem[\protect\citeauthoryear{Franzen, Vernstrom, Jackson, Hurley-Walker,
  Ekers, Heald, Seymour  \& White}{Franzen et~al.}{2019}]{franzen2019}
Franzen T. M.~O.,  Vernstrom T.,  Jackson C.~A.,  Hurley-Walker N.,  Ekers
  R.~D.,  Heald G.,  Seymour N.,   White S.~V.,  2019, \mn@doi [PASA]
  {10.1017/pasa.2018.52}, 36, e004

\bibitem[\protect\citeauthoryear{Galvin et~al.,}{Galvin et~al.}{2020}]{pink}
Galvin T.~J.,  et~al., 2020, \mn@doi [Monthly Notices of the Royal Astronomical
  Society] {10.1093/mnras/staa1890}, 497, 2730

\bibitem[\protect\citeauthoryear{Gilli et~al.,}{Gilli et~al.}{2005}]{gilli05}
Gilli R.,  et~al., 2005, \mn@doi [A\&A] {10.1051/0004-6361:20041375}, 430, 811

\bibitem[\protect\citeauthoryear{Gilli et~al.,}{Gilli et~al.}{2007}]{gilli07}
Gilli R.,  et~al., 2007, \mn@doi [A\&A] {10.1051/0004-6361:20077506}, 475, 83

\bibitem[\protect\citeauthoryear{Gilli et~al.,}{Gilli et~al.}{2009}]{gilli}
Gilli R.,  et~al., 2009, \mn@doi [A\&A] {10.1051/0004-6361:200810821}, 494, 33

\bibitem[\protect\citeauthoryear{Gürkan et~al.,}{Gürkan
  et~al.}{2018}]{Gurkan2018}
Gürkan G.,  et~al., 2018, \mn@doi [Monthly Notices of the Royal Astronomical
  Society] {10.1093/mnras/sty016}, 475, 3010

\bibitem[\protect\citeauthoryear{Hale, Jarvis, Delvecchio, Hatfield, Novak,
  Smolčić  \& Zamorani}{Hale et~al.}{2018}]{hale_cosmos}
Hale C.~L.,  Jarvis M.~J.,  Delvecchio I.,  Hatfield P.~W.,  Novak M.,
  Smolčić V.,   Zamorani G.,  2018, \mn@doi [Monthly Notices of the Royal
  Astronomical Society] {10.1093/mnras/stx2954}, 474, 4133

\bibitem[\protect\citeauthoryear{Hale et~al.,}{Hale et~al.}{2019}]{Hale19}
Hale C.~L.,  et~al., 2019, \mn@doi [A\&A] {10.1051/0004-6361/201833906}, 622,
  A4

\bibitem[\protect\citeauthoryear{Hamilton}{Hamilton}{1993}]{Hamilton1993}
Hamilton A. J.~S.,  1993, \mn@doi [The Astrophysical Journal] {10.1086/173288},
  417, 19

\bibitem[\protect\citeauthoryear{Hao, Kennicutt, Johnson, Calzetti, Dale  \&
  Moustakas}{Hao et~al.}{2011}]{Hao2011}
Hao C.-N.,  Kennicutt R.~C.,  Johnson B.~D.,  Calzetti D.,  Dale D.~A.,
  Moustakas J.,  2011, \mn@doi [The Astrophysical Journal]
  {10.1088/0004-637x/741/2/124}, 741, 124

\bibitem[\protect\citeauthoryear{Hardcastle et~al.,}{Hardcastle
  et~al.}{2016}]{hardcastle2016}
Hardcastle M.~J.,  et~al., 2016, \mn@doi [Monthly Notices of the Royal
  Astronomical Society] {10.1093/mnras/stw1763}, 462, 1910

\bibitem[\protect\citeauthoryear{Heinis et~al.,}{Heinis
  et~al.}{2009}]{Heinis2009}
Heinis S.,  et~al., 2009, \mn@doi [The Astrophysical Journal]
  {10.1088/0004-637x/698/2/1838}, 698, 1838

\bibitem[\protect\citeauthoryear{Hildebrandt et~al.,}{Hildebrandt
  et~al.}{2016}]{rcslens}
Hildebrandt H.,  et~al., 2016, \mn@doi [Monthly Notices of the Royal
  Astronomical Society] {10.1093/mnras/stw2013}, 463, 635

\bibitem[\protect\citeauthoryear{Huynh, Jackson, Norris  \& Prandoni}{Huynh
  et~al.}{2005}]{Huynh_2005}
Huynh M.~T.,  Jackson C.~A.,  Norris R.~P.,   Prandoni I.,  2005, \mn@doi [The
  Astronomical Journal] {10.1086/432873}, 130, 1373

\bibitem[\protect\citeauthoryear{Ineson, Croston, Hardcastle, Kraft, Evans  \&
  Jarvis}{Ineson et~al.}{2015}]{ineson}
Ineson J.,  Croston J.~H.,  Hardcastle M.~J.,  Kraft R.~P.,  Evans D.~A.,
  Jarvis M.,  2015, \mn@doi [Monthly Notices of the Royal Astronomical Society]
  {10.1093/mnras/stv1807}, 453, 2682

\bibitem[\protect\citeauthoryear{Intema}{Intema}{2014}]{Intema2014}
Intema H.~T.,  2014, in Astronomical Society of India Conference Series.
  p.~469, \url {https://ui.adsabs.harvard.edu/abs/2014ASInC..13..469I}

\bibitem[\protect\citeauthoryear{Intema, van~der Tol, Cotton, Cohen, van Bemmel
   \& R\"ottgering}{Intema et~al.}{2009}]{Intema2009}
Intema H.~T.,  van~der Tol S.,  Cotton W.~D.,  Cohen A.~S.,  van Bemmel I.~M.,
   R\"ottgering H. J.~A.,  2009, \mn@doi [A\&A] {10.1051/0004-6361/200811094},
  501, 1185

\bibitem[\protect\citeauthoryear{Intema, van Weeren, R\"ottgering  \&
  Lal}{Intema et~al.}{2011}]{intema2011}
Intema H.~T.,  van Weeren R.~J.,  R\"ottgering H. J.~A.,   Lal D.~V.,  2011,
  \mn@doi [A\&A] {10.1051/0004-6361/201014253}, 535, A38

\bibitem[\protect\citeauthoryear{Intema, Jagannathan, Mooley  \& Frail}{Intema
  et~al.}{2017}]{Intema16}
Intema H.~T.,  Jagannathan P.,  Mooley K.~P.,   Frail D.~A.,  2017, \mn@doi
  [A\&A] {10.1051/0004-6361/201628536}, 598, A78

\bibitem[\protect\citeauthoryear{Jarvis, Rawlings, Eales, Blundell, Bunker,
  Croft, McLure  \& Willott}{Jarvis et~al.}{2001}]{jarvis2001a}
Jarvis M.~J.,  Rawlings S.,  Eales S.,  Blundell K.~M.,  Bunker A.~J.,  Croft
  S.,  McLure R.~J.,   Willott C.~J.,  2001, \mn@doi [Monthly Notices of the
  Royal Astronomical Society] {10.1111/j.1365-2966.2001.04730.x}, 326, 1585

\bibitem[\protect\citeauthoryear{Jarvis, Bernstein  \& Jain}{Jarvis
  et~al.}{2004}]{treecorr}
Jarvis M.,  Bernstein G.,   Jain B.,  2004, \mn@doi [Monthly Notices of the
  Royal Astronomical Society] {10.1111/j.1365-2966.2004.07926.x}, 352, 338

\bibitem[\protect\citeauthoryear{Jarvis et~al.,}{Jarvis
  et~al.}{2010}]{Jarvis2010}
Jarvis M.~J.,  et~al., 2010, \mn@doi [Monthly Notices of the Royal Astronomical
  Society] {10.1111/j.1365-2966.2010.17772.x}, 409, 92

\bibitem[\protect\citeauthoryear{Jarvis et~al.,}{Jarvis
  et~al.}{2016}]{Jarvis2016}
Jarvis M.,  et~al., 2016, in MeerKAT Science: On the Pathway to the SKA. p.~6,
  \url {https://ui.adsabs.harvard.edu/abs/2016mks..confE...6J}

\bibitem[\protect\citeauthoryear{Kaiser}{Kaiser}{1984}]{Kaiser1984}
Kaiser N.,  1984, \mn@doi [The Astrophysical Journal] {10.1086/184341}, 284, L9

\bibitem[\protect\citeauthoryear{Kerscher, Szapudi  \& Szalay}{Kerscher
  et~al.}{2000}]{Kerscher_2000}
Kerscher M.,  Szapudi I.,   Szalay A.~S.,  2000, \mn@doi [The Astrophysical
  Journal] {10.1086/312702}, 535, L13

\bibitem[\protect\citeauthoryear{Lacey \& Cole}{Lacey \&
  Cole}{1993}]{lacey_cole1993}
Lacey C.,  Cole S.,  1993, \mn@doi [Monthly Notices of the Royal Astronomical
  Society] {10.1093/mnras/262.3.627}, 262, 627

\bibitem[\protect\citeauthoryear{Lacey \& Cole}{Lacey \&
  Cole}{1994}]{lacey_cole_1994}
Lacey C.,  Cole S.,  1994, \mn@doi [Monthly Notices of the Royal Astronomical
  Society] {10.1093/mnras/271.3.676}, 271, 676

\bibitem[\protect\citeauthoryear{Landy \& Szalay}{Landy \&
  Szalay}{1993}]{Landy1993}
Landy S.~D.,  Szalay A.~S.,  1993, \mn@doi [The Astrophysical Journal]
  {10.1086/172900}, 412, 64

\bibitem[\protect\citeauthoryear{Lawrence et~al.,}{Lawrence
  et~al.}{2007}]{ukidss}
Lawrence A.,  et~al., 2007, \mn@doi [Monthly Notices of the Royal Astronomical
  Society] {10.1111/j.1365-2966.2007.12040.x}, 379, 1599

\bibitem[\protect\citeauthoryear{Lewis, Bunclark, Irwin, McMahon  \&
  Walton}{Lewis et~al.}{2000}]{int}
Lewis J.~R.,  Bunclark P.~S.,  Irwin M.~J.,  McMahon R.~G.,   Walton N.~A.,
  2000, in Astronomical Data Analysis Software and Systems IX. p.~415, \url
  {https://ui.adsabs.harvard.edu/abs/2000ASPC..216..415L}

\bibitem[\protect\citeauthoryear{Limber}{Limber}{1953}]{Limber1953}
Limber D.~N.,  1953, \mn@doi [The Astrophysical Journal] {10.1086/145672}, 117,
  134

\bibitem[\protect\citeauthoryear{Lindsay et~al.,}{Lindsay
  et~al.}{2014}]{lindsay_first}
Lindsay S.~N.,  et~al., 2014, \mn@doi [Monthly Notices of the Royal
  Astronomical Society] {10.1093/mnras/stu354}, 440, 1527

\bibitem[\protect\citeauthoryear{Ling, Frenk  \& Barrow}{Ling
  et~al.}{1986}]{bootstrap}
Ling E.~N.,  Frenk C.~S.,   Barrow J.~D.,  1986, \mn@doi [Monthly Notices of
  the Royal Astronomical Society] {10.1093/mnras/223.1.21P}, 223, 21P

\bibitem[\protect\citeauthoryear{Magliocchetti, Maddox, Lahav  \&
  Wall}{Magliocchetti et~al.}{1998}]{maglio98}
Magliocchetti M.,  Maddox S.~J.,  Lahav O.,   Wall J.~V.,  1998, \mn@doi
  [Monthly Notices of the Royal Astronomical Society]
  {10.1046/j.1365-8711.1998.01904.x}, 300, 257

\bibitem[\protect\citeauthoryear{Magliocchetti et~al.,}{Magliocchetti
  et~al.}{2002}]{Maglio2002}
Magliocchetti M.,  et~al., 2002, \mn@doi [Monthly Notices of the Royal
  Astronomical Society] {10.1046/j.1365-8711.2002.05386.x}, 333, 100

\bibitem[\protect\citeauthoryear{Magliocchetti et~al.,}{Magliocchetti
  et~al.}{2004}]{Maglio2004}
Magliocchetti M.,  et~al., 2004, \mn@doi [Monthly Notices of the Royal
  Astronomical Society] {10.1111/j.1365-2966.2004.07751.x}, 350, 1485

\bibitem[\protect\citeauthoryear{Magliocchetti et~al.,}{Magliocchetti
  et~al.}{2013}]{maglio2013}
Magliocchetti M.,  et~al., 2013, \mn@doi [Monthly Notices of the Royal
  Astronomical Society] {10.1093/mnras/stt708}, 433, 127

\bibitem[\protect\citeauthoryear{Magliocchetti et~al.,}{Magliocchetti
  et~al.}{2014}]{maglio2014}
Magliocchetti M.,  et~al., 2014, \mn@doi [Monthly Notices of the Royal
  Astronomical Society] {10.1093/mnras/stu863}, 442, 682

\bibitem[\protect\citeauthoryear{Magliocchetti, Lutz, Santini, Salvato,
  Popesso, Berta  \& Pozzi}{Magliocchetti et~al.}{2015}]{maglio2016}
Magliocchetti M.,  Lutz D.,  Santini P.,  Salvato M.,  Popesso P.,  Berta S.,
  Pozzi F.,  2015, \mn@doi [Monthly Notices of the Royal Astronomical Society]
  {10.1093/mnras/stv2645}, 456, 431

\bibitem[\protect\citeauthoryear{Magliocchetti, Popesso, Brusa, Salvato,
  Laigle, McCracken  \& Ilbert}{Magliocchetti et~al.}{2017}]{maglio2017}
Magliocchetti M.,  Popesso P.,  Brusa M.,  Salvato M.,  Laigle C.,  McCracken
  H.~J.,   Ilbert O.,  2017, \mn@doi [Monthly Notices of the Royal Astronomical
  Society] {10.1093/mnras/stw2541}, 464, 3271

\bibitem[\protect\citeauthoryear{Magnelli et~al.,}{Magnelli
  et~al.}{2015}]{magnelli}
Magnelli B.,  et~al., 2015, \mn@doi [A\&A] {10.1051/0004-6361/201424937}, 573,
  A45

\bibitem[\protect\citeauthoryear{Matteo, Perna, Abel  \& Rees}{Matteo
  et~al.}{2002}]{DiMatteo2002}
Matteo T.~D.,  Perna R.,  Abel T.,   Rees M.~J.,  2002, \mn@doi [ApJ]
  {10.1086/324293}, 564, 576

\bibitem[\protect\citeauthoryear{Mauduit et~al.,}{Mauduit et~al.}{2012}]{servs}
Mauduit J.-C.,  et~al., 2012, \mn@doi [Publications of the Astronomical Society
  of the Pacific] {10.1086/666945}, 124, 714

\bibitem[\protect\citeauthoryear{Mazumder, Chakraborty, Datta, Choudhuri, Roy,
  Wadadekar  \& Ishwara-Chandra}{Mazumder et~al.}{2020}]{aishrila1}
Mazumder A.,  Chakraborty A.,  Datta A.,  Choudhuri S.,  Roy N.,  Wadadekar Y.,
    Ishwara-Chandra C.~H.,  2020, \mn@doi [Monthly Notices of the Royal
  Astronomical Society] {10.1093/mnras/staa1317}, 495, 4071

\bibitem[\protect\citeauthoryear{McAlpine, Jarvis  \& Bonfield}{McAlpine
  et~al.}{2013}]{mcalpine2013}
McAlpine K.,  Jarvis M.~J.,   Bonfield D.~G.,  2013, \mn@doi [Monthly Notices
  of the Royal Astronomical Society] {10.1093/mnras/stt1638}, 436, 1084

\bibitem[\protect\citeauthoryear{Mendez et~al.,}{Mendez
  et~al.}{2016}]{Mendez_2016}
Mendez A.~J.,  et~al., 2016, \mn@doi [The Astrophysical Journal]
  {10.3847/0004-637x/821/1/55}, 821, 55

\bibitem[\protect\citeauthoryear{{Mignano, A.}, {Prandoni, I.}, {Gregorini,
  L.}, {Parma, P.}, {de Ruiter, H. R.}, {Wieringa, M. H.}, {Vettolani, G.}  \&
  {Ekers, R. D.}}{{Mignano, A.} et~al.}{2008}]{mignano}
{Mignano, A.} {Prandoni, I.} {Gregorini, L.} {Parma, P.} {de Ruiter, H. R.}
  {Wieringa, M. H.} {Vettolani, G.}  {Ekers, R. D.} 2008, \mn@doi [A\&A]
  {10.1051/0004-6361:20078545}, 477, 459

\bibitem[\protect\citeauthoryear{Mo, van~den Bosch  \& White}{Mo
  et~al.}{2010}]{Mo2010}
Mo H.,  van~den Bosch F.,   White S.,  2010, Galaxy Formation and Evolution.
Cambridge University Press, Cambridge, \url
  {https://www.cambridge.org/core/books/galaxy-formation-and-evolution/E236D9F26B797202BCA28637BF17E75F}

\bibitem[\protect\citeauthoryear{{Mohan} \& {Rafferty}}{{Mohan} \&
  {Rafferty}}{2015}]{Mohan2015}
{Mohan} N.,  {Rafferty} D.,  2015, {PyBDSF: Python Blob Detection and Source
  Finder} (\mn@eprint {ascl} {1502.007})

\bibitem[\protect\citeauthoryear{Muzzin, Wilson  \& Collaboration}{Muzzin
  et~al.}{2007}]{sparcs}
Muzzin A.,  Wilson G.,   Collaboration S.,  2007, in American Astronomical
  Society Meeting Abstracts. p. 78.05, \url
  {https://ui.adsabs.harvard.edu/abs/2007AAS...211.7805M}

\bibitem[\protect\citeauthoryear{Norberg et~al.,}{Norberg
  et~al.}{2001}]{norberg}
Norberg P.,  et~al., 2001, \mn@doi [Monthly Notices of the Royal Astronomical
  Society] {10.1046/j.1365-8711.2001.04839.x}, 328, 64

\bibitem[\protect\citeauthoryear{Norris et~al.,}{Norris
  et~al.}{2011}]{Norris2011}
Norris R.~P.,  et~al., 2011, \mn@doi [Publications of the Astronomical Society
  of Australia] {10.1071/AS11021}, 28, 215

\bibitem[\protect\citeauthoryear{Norris et~al.,}{Norris
  et~al.}{2013}]{Norris2013}
Norris R.~P.,  et~al., 2013, \mn@doi [Publications of the Astronomical Society
  of Australia] {10.1017/pas.2012.020}, 30, e020

\bibitem[\protect\citeauthoryear{Norris et~al.,}{Norris
  et~al.}{2021}]{norris2021}
Norris R.~P.,  et~al., 2021, \mn@doi [Publications of the Astronomical Society
  of Australia] {10.1017/pasa.2021.42}, 38, e046

\bibitem[\protect\citeauthoryear{Ocran, Taylor, Vaccari, Ishwara-Chandra,
  Prandoni, Prescott  \& Mancuso}{Ocran et~al.}{2019}]{ocran_sfg}
Ocran E.~F.,  Taylor A.~R.,  Vaccari M.,  Ishwara-Chandra C.~H.,  Prandoni I.,
  Prescott M.,   Mancuso C.,  2019, \mn@doi [Monthly Notices of the Royal
  Astronomical Society] {10.1093/mnras/stz3401}, 491, 5911

\bibitem[\protect\citeauthoryear{Ocran, Taylor, Vaccari, Ishwara-Chandra,
  Prandoni, Prescott  \& Mancuso}{Ocran et~al.}{2020}]{ocran_agn}
Ocran E.~F.,  Taylor A.~R.,  Vaccari M.,  Ishwara-Chandra C.~H.,  Prandoni I.,
  Prescott M.,   Mancuso C.,  2020, \mn@doi [Monthly Notices of the Royal
  Astronomical Society] {10.1093/mnras/staa3538}, 500, 4685

\bibitem[\protect\citeauthoryear{Oort}{Oort}{1987}]{Oort1987}
Oort M. J.~A.,  1987, PhD thesis, Leiden Observatory, \url
  {https://ui.adsabs.harvard.edu/abs/1987PhDT........40O}

\bibitem[\protect\citeauthoryear{Overzier, R\"ottgering, Rengelink  \&
  Wilman}{Overzier et~al.}{2003}]{overzier2003}
Overzier R.~A.,  R\"ottgering H. J.~A.,  Rengelink R.~B.,   Wilman R.~J.,
  2003, \mn@doi [A\&A] {10.1051/0004-6361:20030527}, 405, 53

\bibitem[\protect\citeauthoryear{Padovani}{Padovani}{2016}]{Padovani2016}
Padovani P.,  2016, \mn@doi [The Astronomy and Astrophysics Review]
  {10.1007/s00159-016-0098-6}, 24, 13

\bibitem[\protect\citeauthoryear{Padovani, Bonzini, Kellermann, Miller,
  Mainieri  \& Tozzi}{Padovani et~al.}{2015}]{padovani2015}
Padovani P.,  Bonzini M.,  Kellermann K.~I.,  Miller N.,  Mainieri V.,   Tozzi
  P.,  2015, \mn@doi [Monthly Notices of the Royal Astronomical Society]
  {10.1093/mnras/stv1375}, 452, 1263

\bibitem[\protect\citeauthoryear{Peacock \& Smith}{Peacock \&
  Smith}{2000}]{peacock}
Peacock J.~A.,  Smith R.~E.,  2000, \mn@doi [Monthly Notices of the Royal
  Astronomical Society] {10.1046/j.1365-8711.2000.03779.x}, 318, 1144

\bibitem[\protect\citeauthoryear{Peacock et~al.,}{Peacock
  et~al.}{2001}]{Peacock2001}
Peacock J.~A.,  et~al., 2001, Nature, 410, 169

\bibitem[\protect\citeauthoryear{Peebles}{Peebles}{1980}]{Peebles1980}
Peebles P. J.~E.,  1980, The large-scale structure of the universe.
Princeton University Press

\bibitem[\protect\citeauthoryear{Percival et~al.,}{Percival
  et~al.}{2001}]{percival}
Percival W.~J.,  et~al., 2001, \mn@doi [Monthly Notices of the Royal
  Astronomical Society] {10.1046/j.1365-8711.2001.04827.x}, 327, 1297

\bibitem[\protect\citeauthoryear{{Planck Collaboration} et~al.,}{{Planck
  Collaboration} et~al.}{2014}]{planck2013}
{Planck Collaboration} et~al., 2014, \mn@doi [A\&A]
  {10.1051/0004-6361/201321543}, 571, A17

\bibitem[\protect\citeauthoryear{{Planck Collaboration} et~al.,}{{Planck
  Collaboration} et~al.}{2019}]{Planck2018I}
{Planck Collaboration} et~al., 2019, \mn@doi [A\&A]
  {10.1051/0004-6361/201833880}

\bibitem[\protect\citeauthoryear{Prandoni, Guglielmino, Morganti, Vaccari,
  Maini, Röttgering, Jarvis  \& Garrett}{Prandoni et~al.}{2018}]{prandoni2018}
Prandoni I.,  Guglielmino G.,  Morganti R.,  Vaccari M.,  Maini A.,
  Röttgering H. J.~A.,  Jarvis M.~J.,   Garrett M.~A.,  2018, \mn@doi [MNRAS]
  {10.1093/mnras/sty2521}, 481, 4548

\bibitem[\protect\citeauthoryear{Prescott et~al.,}{Prescott
  et~al.}{2016}]{prescott_gama}
Prescott M.,  et~al., 2016, \mn@doi [Monthly Notices of the Royal Astronomical
  Society] {10.1093/mnras/stv3020}, 457, 730

\bibitem[\protect\citeauthoryear{Press \& Schechter}{Press \&
  Schechter}{1974}]{PressSchechter1974}
Press W.~H.,  Schechter P.,  1974, \mn@doi [The Astrophysical Journal]
  {10.1086/152650}, 187, 425

\bibitem[\protect\citeauthoryear{{Price-Whelan} et~al.,}{{Price-Whelan}
  et~al.}{2018}]{astropy:2018}
{Price-Whelan} A.~M.,  et~al., 2018, \mn@doi [\aj] {10.3847/1538-3881/aabc4f},
  \href {https://ui.adsabs.harvard.edu/#abs/2018AJ....156..123T} {156, 123}

\bibitem[\protect\citeauthoryear{Raccanelli et~al.,}{Raccanelli
  et~al.}{2012}]{raccanelli2012}
Raccanelli A.,  et~al., 2012, \mn@doi [Monthly Notices of the Royal
  Astronomical Society] {10.1111/j.1365-2966.2012.20634.x}, 424, 801

\bibitem[\protect\citeauthoryear{Raccanelli et~al.,}{Raccanelli
  et~al.}{2015}]{Raccanelli_2015}
Raccanelli A.,  et~al., 2015, \mn@doi [Journal of Cosmology and Astroparticle
  Physics] {10.1088/1475-7516/2015/01/042}, 2015, 042

\bibitem[\protect\citeauthoryear{Rana \& Bagla}{Rana \&
  Bagla}{2019}]{rana_tgss}
Rana S.,  Bagla J.~S.,  2019, \mn@doi [Monthly Notices of the Royal
  Astronomical Society] {10.1093/mnras/stz831}, 485, 5891

\bibitem[\protect\citeauthoryear{Ross et~al.,}{Ross et~al.}{2007}]{2df_spatial}
Ross N.~P.,  et~al., 2007, \mn@doi [Monthly Notices of the Royal Astronomical
  Society] {10.1111/j.1365-2966.2007.12289.x}, 381, 573

\bibitem[\protect\citeauthoryear{Rowan-Robinson et~al.,}{Rowan-Robinson
  et~al.}{2008}]{robinson2008}
Rowan-Robinson M.,  et~al., 2008, \mn@doi [Monthly Notices of the Royal
  Astronomical Society] {10.1111/j.1365-2966.2008.13109.x}, 386, 697

\bibitem[\protect\citeauthoryear{Rowan-Robinson, Gonzalez-Solares, Vaccari  \&
  Marchetti}{Rowan-Robinson et~al.}{2012}]{robinson2012}
Rowan-Robinson M.,  Gonzalez-Solares E.,  Vaccari M.,   Marchetti L.,  2012,
  \mn@doi [Monthly Notices of the Royal Astronomical Society]
  {10.1093/mnras/sts163}, 428, 1958

\bibitem[\protect\citeauthoryear{{Sabater, J.} et~al.,}{{Sabater, J.}
  et~al.}{2021}]{sabater}
{Sabater, J.} et~al., 2021, \mn@doi [A\&A] {10.1051/0004-6361/202038828}, 648,
  A2

\bibitem[\protect\citeauthoryear{Salazar-Albornoz et~al.,}{Salazar-Albornoz
  et~al.}{2017}]{boss_tomography}
Salazar-Albornoz S.,  et~al., 2017, \mn@doi [Monthly Notices of the Royal
  Astronomical Society] {10.1093/mnras/stx633}, 468, 2938

\bibitem[\protect\citeauthoryear{Saxena, Röttgering  \& Rigby}{Saxena
  et~al.}{2017}]{saxena}
Saxena A.,  Röttgering H. J.~A.,   Rigby E.~E.,  2017, \mn@doi [Monthly
  Notices of the Royal Astronomical Society] {10.1093/mnras/stx1150}, 469, 4083

\bibitem[\protect\citeauthoryear{Seldner \& Peebles}{Seldner \&
  Peebles}{1981}]{seldner1981}
Seldner M.,  Peebles P. J.~E.,  1981, \mn@doi [Monthly Notices of the Royal
  Astronomical Society] {10.1093/mnras/194.2.251}, 194, 251

\bibitem[\protect\citeauthoryear{Seljak}{Seljak}{2009}]{seljak}
Seljak U. c.~v.,  2009, \mn@doi [Phys. Rev. Lett.]
  {10.1103/PhysRevLett.102.021302}, 102, 021302

\bibitem[\protect\citeauthoryear{Seymour et~al.,}{Seymour
  et~al.}{2008}]{seymour}
Seymour N.,  et~al., 2008, \mn@doi [Monthly Notices of the Royal Astronomical
  Society] {10.1111/j.1365-2966.2008.13166.x}, 386, 1695

\bibitem[\protect\citeauthoryear{Shaver \& Pierre}{Shaver \&
  Pierre}{1989}]{Shaver1989}
Shaver P.~A.,  Pierre M.,  1989, Astronomy and Astrophysics, 220, 35

\bibitem[\protect\citeauthoryear{Shaver, Windhorst, Madau  \& de Bruyn}{Shaver
  et~al.}{1999}]{Shaver1999}
Shaver P.~A.,  Windhorst R.~A.,  Madau P.,   de Bruyn A.~G.,  1999, A\&A, 345,
  380

\bibitem[\protect\citeauthoryear{Sheth \& Tormen}{Sheth \&
  Tormen}{1999}]{sheth_tormen}
Sheth R.~K.,  Tormen G.,  1999, \mn@doi [Monthly Notices of the Royal
  Astronomical Society] {10.1046/j.1365-8711.1999.02692.x}, 308, 119

\bibitem[\protect\citeauthoryear{Shi et~al.,}{Shi et~al.}{2016}]{Shi2016}
Shi F.,  et~al., 2016, \mn@doi [The Astrophysical Journal]
  {10.3847/1538-4357/833/2/241}, 833, 241

\bibitem[\protect\citeauthoryear{{Shimwell, T. W.} et~al.,}{{Shimwell, T. W.}
  et~al.}{2017}]{lotss_dr1}
{Shimwell, T. W.} et~al., 2017, \mn@doi [A\&A] {10.1051/0004-6361/201629313},
  598, A104

\bibitem[\protect\citeauthoryear{{Shimwell, T. W.} et~al.,}{{Shimwell, T. W.}
  et~al.}{2019}]{lotss2019}
{Shimwell, T. W.} et~al., 2019, \mn@doi [A\&A] {10.1051/0004-6361/201833559},
  622, A1

\bibitem[\protect\citeauthoryear{Shirley et~al.,}{Shirley et~al.}{2019}]{help1}
Shirley R.,  et~al., 2019, \mn@doi [Monthly Notices of the Royal Astronomical
  Society] {10.1093/mnras/stz2509}, 490, 634

\bibitem[\protect\citeauthoryear{Siewert et~al.,}{Siewert
  et~al.}{2020}]{lotss_clustering}
Siewert T.~M.,  et~al., 2020, \mn@doi [A\&A] {10.1051/0004-6361/201936592},
  643, A100

\bibitem[\protect\citeauthoryear{Simpson et~al.,}{Simpson
  et~al.}{2006}]{simpson2006}
Simpson C.,  et~al., 2006, \mn@doi [Monthly Notices of the Royal Astronomical
  Society] {10.1111/j.1365-2966.2006.10907.x}, 372, 741

\bibitem[\protect\citeauthoryear{Singh et~al.,}{Singh et~al.}{2014}]{highz}
Singh V.,  et~al., 2014, \mn@doi [A\&A] {10.1051/0004-6361/201423644}, 569, A52

\bibitem[\protect\citeauthoryear{Smol{\v{c}}i{\'{c}}
  et~al.,}{Smol{\v{c}}i{\'{c}} et~al.}{2008}]{Smolic2008}
Smol{\v{c}}i{\'{c}} V.,  et~al., 2008, \mn@doi [The Astrophysical Journal
  Supplement Series] {10.1086/588028}, 177, 14

\bibitem[\protect\citeauthoryear{Smolci\'{}c et~al.,}{Smolci\'{}c
  et~al.}{2017}]{smolic_agn}
Smolci\'{}c V.,  et~al., 2017, \mn@doi [A\&A] {10.1051/0004-6361/201730685},
  602, A6

\bibitem[\protect\citeauthoryear{Starikova, Berta, Franceschini, Marchetti,
  Rodighiero, Vaccari  \& Vikhlinin}{Starikova et~al.}{2012}]{Starikova_2012}
Starikova S.,  Berta S.,  Franceschini A.,  Marchetti L.,  Rodighiero G.,
  Vaccari M.,   Vikhlinin A.,  2012, \mn@doi [The Astrophysical Journal]
  {10.1088/0004-637x/751/2/126}, 751, 126

\bibitem[\protect\citeauthoryear{Swarup, Ananthakrishnan, Kapahi, Rao,
  Subrahmanya  \& Kulkarni}{Swarup et~al.}{1991}]{Swarup1991}
Swarup G.,  Ananthakrishnan S.,  Kapahi V.~K.,  Rao A.~P.,  Subrahmanya C.~R.,
   Kulkarni V.~K.,  1991, Current Science, Vol. 60, NO.2/JAN25, P. 95, 1991,
  60, 95

\bibitem[\protect\citeauthoryear{Szokoly et~al.,}{Szokoly
  et~al.}{2004}]{Szokoly_2004}
Szokoly G.~P.,  et~al., 2004, \mn@doi [The Astrophysical Journal Supplement
  Series] {10.1086/424707}, 155, 271

\bibitem[\protect\citeauthoryear{Tasse et~al.,}{Tasse et~al.}{2021}]{tasse2020}
Tasse C.,  et~al., 2021, \mn@doi [A\&A] {10.1051/0004-6361/202038804}, 648, A1

\bibitem[\protect\citeauthoryear{Tiwari, Ghosh  \& Jain}{Tiwari
  et~al.}{2019}]{Tiwari_2019}
Tiwari P.,  Ghosh S.,   Jain P.,  2019, \mn@doi [The Astrophysical Journal]
  {10.3847/1538-4357/ab54c8}, 887, 175

\bibitem[\protect\citeauthoryear{Trott et~al.,}{Trott
  et~al.}{2016}]{Trott_2016}
Trott C.~M.,  et~al., 2016, \mn@doi [ApJ] {10.3847/0004-637x/818/2/139}, 818,
  139

\bibitem[\protect\citeauthoryear{Vernstrom, Scott, Wall, Condon, Cotton,
  Kellermann  \& Perley}{Vernstrom et~al.}{2016}]{vernstrom2016}
Vernstrom T.,  Scott D.,  Wall J.~V.,  Condon J.~J.,  Cotton W.~D.,  Kellermann
  K.~I.,   Perley R.~A.,  2016, \mn@doi [Monthly Notices of the Royal
  Astronomical Society] {10.1093/mnras/stw1836}, 462, 2934

\bibitem[\protect\citeauthoryear{Wang, Brunner  \& Dolence}{Wang
  et~al.}{2013}]{sdss_wang}
Wang Y.,  Brunner R.~J.,   Dolence J.~C.,  2013, \mn@doi [Monthly Notices of
  the Royal Astronomical Society] {10.1093/mnras/stt450}, 432, 1961

\bibitem[\protect\citeauthoryear{{Williams, W. L.} et~al.,}{{Williams, W. L.}
  et~al.}{2019}]{lofar_association}
{Williams, W. L.} et~al., 2019, \mn@doi [A\&A] {10.1051/0004-6361/201833564},
  622, A2

\bibitem[\protect\citeauthoryear{Williams, Intema  \& R\"ottgering}{Williams
  et~al.}{2013}]{william2013}
Williams W.~L.,  Intema H.~T.,   R\"ottgering H. J.~A.,  2013, \mn@doi [A\&A]
  {10.1051/0004-6361/201220235}, 549, A55

\bibitem[\protect\citeauthoryear{Williams et~al.,}{Williams
  et~al.}{2018}]{williams2018}
Williams W.~L.,  et~al., 2018, \mn@doi [Monthly Notices of the Royal
  Astronomical Society] {10.1093/mnras/sty026}, 475, 3429

\bibitem[\protect\citeauthoryear{Wilman et~al.,}{Wilman et~al.}{2008}]{skads}
Wilman R.~J.,  et~al., 2008, \mn@doi [Monthly Notices of the Royal Astronomical
  Society] {10.1111/j.1365-2966.2008.13486.x}, 388, 1335

\bibitem[\protect\citeauthoryear{York et~al.,}{York et~al.}{2000}]{SDSSI}
York D.~G.,  et~al., 2000, \mn@doi [The Astronomical Journal] {10.1086/301513},
  120, 1579

\bibitem[\protect\citeauthoryear{de Simoni et~al.,}{de~Simoni
  et~al.}{2013}]{sdss_simoni}
de Simoni F.,  et~al., 2013, \mn@doi [Monthly Notices of the Royal Astronomical
  Society] {10.1093/mnras/stt1496}, 435, 3017

\bibitem[\protect\citeauthoryear{{van Haarlem, M. P.} et~al.,}{{van Haarlem, M.
  P.} et~al.}{2013}]{lofar}
{van Haarlem, M. P.} et~al., 2013, \mn@doi [A\&A]
  {10.1051/0004-6361/201220873}, 556, A2

\makeatother
\end{thebibliography}





\bsp	
\label{lastpage}
\end{document}